\documentclass[12pt, manuscript]{aastex}
\pdfoutput=1

\shorttitle{Proper Motions and Shocks in HH 7-11 Stellar Jet}
\shortauthors{Hartigan, Holcomb and Frank}

\begin{document}

\title{Proper Motions and Shock Wave Dynamics in the HH 7-11 Stellar Jet}

 

\author{P. Hartigan and R. Holcomb\altaffilmark{1},
\and A. Frank\altaffilmark{2}}

\altaffiltext{1}{Physics and Astronomy Dept., Rice University, Houston, TX}
\altaffiltext{2}{Physics and Astronomy Dept., University of Rochester, Rochester NY} 

\begin{abstract}
We have used the Hubble Space Telescope to acquire new
broad-band and narrow-band images of the optical line emission and 
red continuum associated with the HH 7-11 stellar jet in the NGC 1333 star formation region.
Combining the new narrow-band images of H$\alpha$, [O~I] $\lambda$6300
and [S~II] $\lambda$6716 allows us to measure electron
densities and excitations at each point in the outflow with the spatial resolution of HST,
while the I-band image traces out the boundary of the cavity evacuated by the outflow.
Comparing these images with those taken $\sim$ 20 years ago yields high precision proper motions
for all the HH objects in the outflow.  HH 11 is a bullet-like clump,
and emerges from the exciting source SVS~13A towards the Earth at
24 degrees to line of sight. In contrast, HH 8 and HH 10 consist of two rings of shocked gas
that show no bulk proper motions even though the emitting gas is blueshifted. The HH~8 rings 
are expanding with time. These shocks mark places where ambient material located along
the path of the jet redirects the outflow.
HH 7 consists of multiple shells, and emits strongly in H$_2$ in what appears to be a terminal bow shock
for the outflow, implying that the jet has yet to fully break out of its nascent cloud core.
The jet largely fragments into clumps by the time it reaches HH~7.
As in the case of HH~110, deflection from ambient material plays a key role in
producing observable shock waves in the HH~7-11 outflow. 

\end{abstract}


\section{Introduction} \label{sec:intro}

Collimated jets arise from accretion disks in a variety of astrophysical systems,
including compact objects and galactic nuclei \citep[e.g.][]{romero17},
and young stars \citep[see][for a review]{frank14}.  Material within an accretion disk loses
angular momentum as it drifts inward onto the central source, and jets play
a primary role in removing this angular momentum from the system in most modern
numerical models \citep[e.g.]{zanni13,nolan17}.
Jets from young stars also affect star formation within 
a dark cloud by reshaping the surrounding medium as the outflows
entrain ambient material and evacuate large cavities within dark clouds.
The dense shells of shocked gas that form at the interfaces between the jet
and the dark cloud deposit energy and momentum into the cloud, and provide
a source of turbulence that opposes the gravitational forces responsible
for creating new stars \citep[e.g.]{frank07}. As jets drive shells and cavities, they contribute
to the complex geometry of clumps and filaments observed
within dark clouds \citep[e.g.][]{herschel}. 

In many ways it is easiest to study collimated outflows in young stars
because shock waves in young stellar jets typically cool radiatively, so the
resulting emission lines make it possible to identify the locations and morphologies
of the shocks, and we can obtain maps of densities and temperatures throughout
the flow. In addition, many young stellar jets exhibit measurable proper
motions and spectral variations over a timespan of several years, so one can
observe how the flows evolve in real time \citep{hartigan11}.  The ability to identify shocks
observationally and follow them over time creates a critical connection with
studies of dynamical instabilities in numerical and laboratory work \citep{frank14, hartigan16}.

In the past two decades, the spatial resolution afforded by HST has transformed
our understanding of jets from young stars. 
Because shocks in stellar jets typically move into neutral gas,
Balmer line emission occurs both in a thin shell excited by collisions at
the shock front as well as within an extended recombination zone if the shock
is strong enough to ionize hydrogen. HST images of jets are able to
resolve the difference in position between the Balmer shell and
the regions where forbidden lines such as [S~II] and [O~I] emit 
\citep{heathcote96}. Such images also provide a means to
identify the direction that the gas flows through the shock.
HST observations of time-variable hot-spots have been identified with
shock intersections \citep{hartigan11,hartigan16}, and there are now
observations of temporal variability in several emission line ratios
\citep{raga16}.  Resolving the size of cooling zones has led to new
theoretical work, allowing Alfv\'enic Mach numbers to be measured in jets \citep{hw15}.
In addition, spatially-resolved determinations of temperature and density maps as well as
precise measures of jet collimation within 50 AU of the star \citep{hm07, deC10}
continue to provide major constraints on disk wind models.
Many outstanding questions remain, including the nature of the magnetic geometry and
collimation close to the source, the roles accretion disks and photoevaporative
flows play in altering the formation of planetesimals, and how the narrow jet
interacts with its surroundings and drives a molecular flow. This last topic is
the main focus of this paper.

Located at a distance of $\sim$ 300~pc \citep[][section \ref{sec:ysos}]{deZ99,belikov02},
the Herbig-Haro objects HH 7-11 were among
the first of their class to be catalogued \citep{herbig74}, and represent one of the first known
examples of a bipolar molecular outflow \citep{se81}.  A bright, variable infrared source
SVS~13A \citep[][sometimes denoted SSV 13]{svs76} 
drives a blueshifted jet in the direction of the HH 7-11 outflow \citep{hodapp14}.
SVS~13A is in fact a close binary separated by only 0.3 arcseconds
\citep[VLA~4A and VLA~4B;][]{anglada00}, and there
is evidence from high-resolution SiO and SO maps that both stars drive outflows
aligned nearly parallel to one another \citep{lef17}.

\begin{figure}
{\centering
\includegraphics[angle=0,scale=1.00]{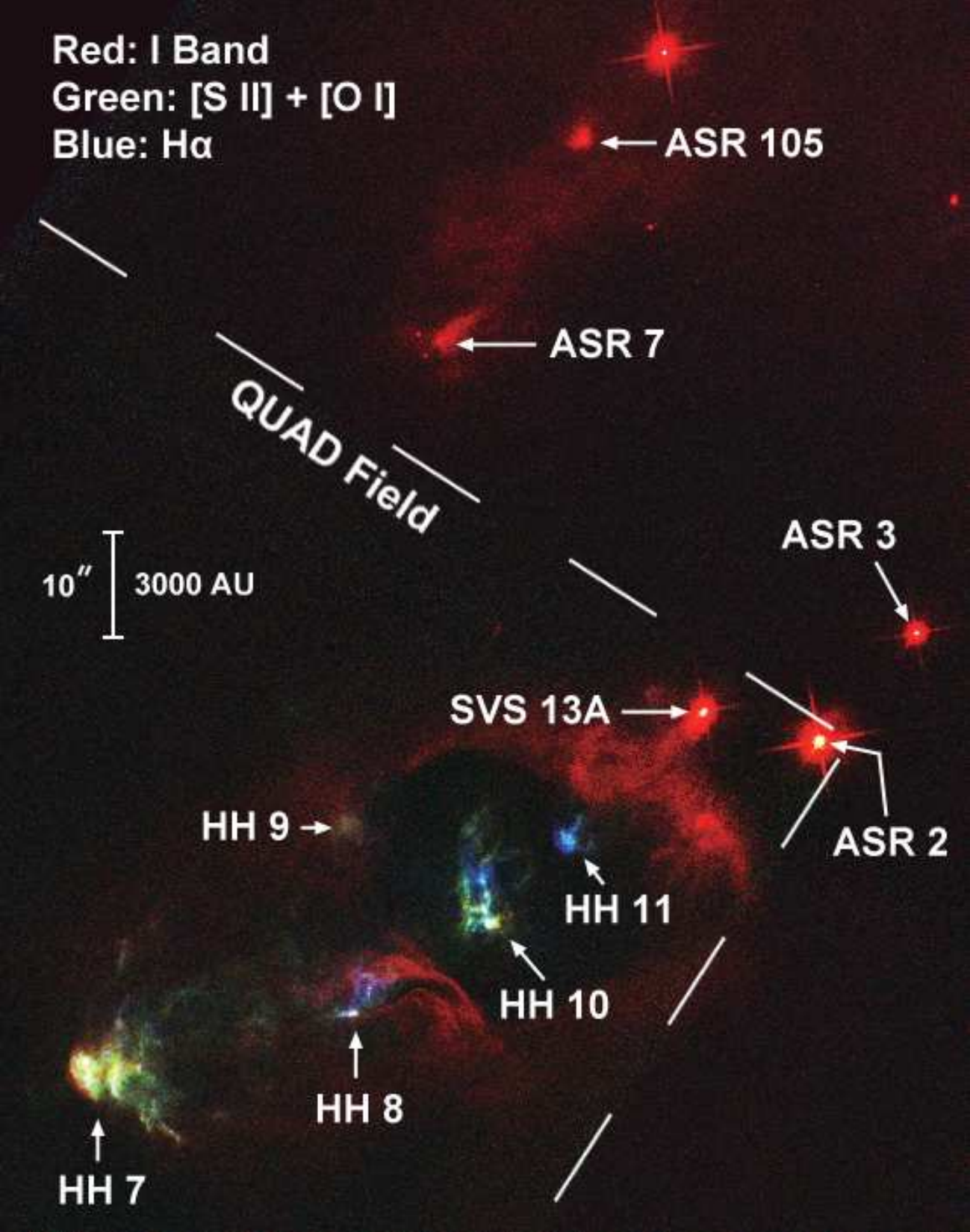}
\caption{Color composite of the HH~7-11 Region from I-band (F850LP,
a Sloan z$^\prime$ filter; red),
[S~II] $\lambda$6716 + [O~I] $\lambda$6300 (FQ672N + F631N; green),
and H$\alpha$ (F656N; blue) HST ACS images. The dashed region marks the boundaries
of the [S~II] QUAD filter. The HH objects and point
sources discussed in the text are labeled. 
SVS~13A is the driving source of the HH 7-11 flow. North is up and east
to the left.
}
\label{fig:overview}
}
\end{figure}

Fig.~\ref{fig:overview} shows the overall morphology of the HH~7-11 outflow.
Proceeding down the HH flow away from the source,
HH 11 appears as a compact knot, HH 8 and 10 are more extended, and HH 7 is a
resolved bow shock.  HH~9 is displaced significantly
to the north of the other HH objects.  HH~11 has a higher blueshifted
radial velocity ($\sim\ -200$ km$\,$s$^{-1}$) compared with HH~7, 8, 9, and~10
($\sim\ -$50 km$\,$s$^{-1}$),
but all the HH objects have modest line widths \citep[$\lesssim$ 100 km$\,$s$^{-1}$;][]{sb87}.
HH~7-11 are unusual compared with other HH knots in that they have very low-excitation
spectra that show strong [O I] and [S II], weak [N II], and no [O III] \citep{bohm83}.
These spectra indicate low shock velocities of $\sim$ 30 km$\,$s$^{-1}$ where the emitting gas is
mostly neutral \citep{hrh87,ds17}. 

Because HH 7, HH 8, and HH 10 also radiate strongly in near-IR H$_2$ quadrupole lines
\citep{garden90,khan03}, these sources
are among the best examples to study shocked cooling zones that possess 
both molecular and optical line emission. High-resolution ground-based
spectra of multiple H$_2$ lines in HH~7 imply a thermal population generally consistent
with a simple C-shocks and inconsistent with fluorescence,
though there is a weak high-temperature component at $\sim$ 5000~K that could arise from
molecular formation on dust \citep{pike16,geballe17}. Molecular lines in HH~7 are also
strong in the mid-IR, and enable studies of ortho/para H$_2$ and HD/H$_2$ ratios
\citep{yuan11,yuan12}.

When combined with emission-line ratio maps and spectral line profiles, proper motions of
spatially-resolved HH shocks create a comprehensive data set that characterizes 
the dynamics of the shocked gas. However, it has been difficult to measure proper motions in
the HH~7-11 outflow because the velocities are low and because a typical field of view
for many instruments includes few stars suitable for alignment. Low-spatial resolution
observations from Spitzer at 4.5$\mu$m suggest proper motions $\lesssim$ 10 km$\,$s$^{-1}$
\citep{raga12}, consistent with revised measurements in H$_2$ that take into account the
photocenter shift of SVS~13A between epochs \citep{khan03}. Existing ground-based
optical proper motions are also low, 
$\sim$ 50 km$\,$s$^{-1}$ directed away from SVS~13A for HH~11,
with smaller values of $\sim$ 30 km$\,$s$^{-1}$ or less indicated for the other HH objects
\citep{hj83,noriega01}. Low proper motions such as these can only be measured accurately
with high spatial-resolution images such as those from HST or with ground-based AO systems.
Recent near-IR AO images of the immediate vicinity of SVS~13A uncovered a remarkable series of nested
cavities along the blueshifted jet, with proper motions measured to be 28 km$\,$s$^{-1}$
\citep[taking the distance to be 300~pc;][]{hodapp14}. These cavities correspond with the bullets of high
velocity molecular gas seen in SiO \citep{lef17}.

In this paper we present new narrowband HST images of the HH~7-11 outflow in
the lines of [O~I] $\lambda$6300, H$\alpha$, and [S~II] $\lambda$6716, together with a narrowband
continuum image at R and a broadband continuum image at I. These are the first optical narrowband
HST images to include all of HH~7, and the first study to make use of the WFC3 Quad filter
on HST to isolate the [S~II] $\lambda$6716 line for stellar jet work. 
The new images allow us to measure high-precision proper motions across the region, and by
combining [S~II] $\lambda$6716 with [O~I] $\lambda$6300 
(and [S~II] $\lambda$6716 with an archival [S~II] $\lambda\lambda$6716+6731 image) we compute
maps of the electron density at the spatial resolution of HST (0.07$^{\prime\prime}$; $\sim$ 20~AU at 300~pc)
across the entire outflow.  We also observe where the shock excitations are highest from
the ratio map of H$\alpha$ / ([O~I] $\lambda$6300 + [S~II] $\lambda$6716).

In what follows,
Sec.~\ref{sec:data} describes the data reduction and analysis procedures, and
Sec.~\ref{sec:ratios} presents the new HST images and ratio maps.
New images of extended young stellar objects in the field, including a close binary
are in Sec.~\ref{sec:ysos}, and Sec.~\ref{sec:motions} derives new proper motion
measurements for the region.  We collect these data together in Sec.~\ref{sec:discussion},
and summarize the paper in Sec.~\ref{sec:summary}. 

\section{Data Reduction} \label{sec:data}

We acquired new images of the HH 7-11 flow  
in [O~I] $\lambda$6300 (F631N; 5779 seconds), H$\alpha$ (F656N; 5778 seconds),
narrowband R-continuum (F645N; 2787 seconds),
broadband I (F850LP; 2725 seconds), and [S~II] $\lambda$6716 (FQ672N; 5778 seconds)
with the WFC3 camera on the Hubble Space Telescope between Dec 11, 2017 and Jan 12, 2018.
The field of view includes the driving source SVS~13A, the entire HH~7-11 outflow,
and several stars to the north and west of SVS~13A. FQ672N is a `quad' filter with
a reduced field of view, but it is possible to fit the entire HH~7-11 outflow
in the field with the proper orientation restrictions (Fig.~\ref{fig:overview}). 
The filters for [O~I] $\lambda$6300, H$\alpha$, and [S~II] $\lambda$6716 are narrow
enough to isolate these lines individually (Sec.~\ref{sec:ratios}),
but broad enough to include the most blueshifted line emission in the flow.
The narrowband continuum filter excludes all emission lines (FWHM 85~\AA ), and 
was used to verify that no continuum was present in the HH objects when calculating
ratio images between emission lines in the R-band.
The I-band filter (a Sloan DSS z$^\prime$), includes a mixture of continuum and emission lines,
and is suitable for showing morphologies of both HH objects and reflected-light cavities.

Image reductions followed the standard WFC3 pipeline procedures,
and produce output images in units of electrons/sec.
To calculate image ratios we need to correct for the different throughputs
in each filter at the wavelength of the emission line, which are
0.232, 0.222, and 0.242, respectively, for [O~I] in F631N, H$\alpha$ 
in F656N, and [S~II] $\lambda$6716 in FQ672N \citep{dressel19}.  The images were aligned to one-another
using the geomap and geotran packages in IRAF, allowing for rotations
and translations between the HST images, and also for a constant scale factor 
in the H$_2$. Typically only $\sim$ 7 stars are in common between the narrowband
HST images owing to high extinction in the region. Only three stars are present
in the quad filter. The rms registration errors
between the narrowband HST images is 0.2 pixels ($\sim$ 0.008 arcseconds).

We supplemented our data with a ground-based H$_2$ image of HH~7-11 taken
Jan 1, 2007 in 0.7 arcsecond seeing as part of the the UKIDSS survey \citep{lucas08}. 
We used the F850LP filter to align with the H$_2$ image because these images
share the most stars, though we had to remove some embedded extended sources
from the fit because extinction from circumstellar disks and envelopes caused
the optical and near-infrared photocenters to differ. Overall,
registration of the H$_2$ image is good to about 0.2 arcseconds.
For reference, motions of 20 km$\,$s$^{-1}$ over the 11-year time interval
between the UKIDSS and HST images amounts to 0.15 arcseconds.
No alignment stars are present in the immediate vicinity of HH~7. 

\section{Difference Images and Emission Line Ratio Maps}
\label{sec:ratios}

\begin{figure}
\centering
\includegraphics[angle=0,scale=1.00,width=\textwidth]{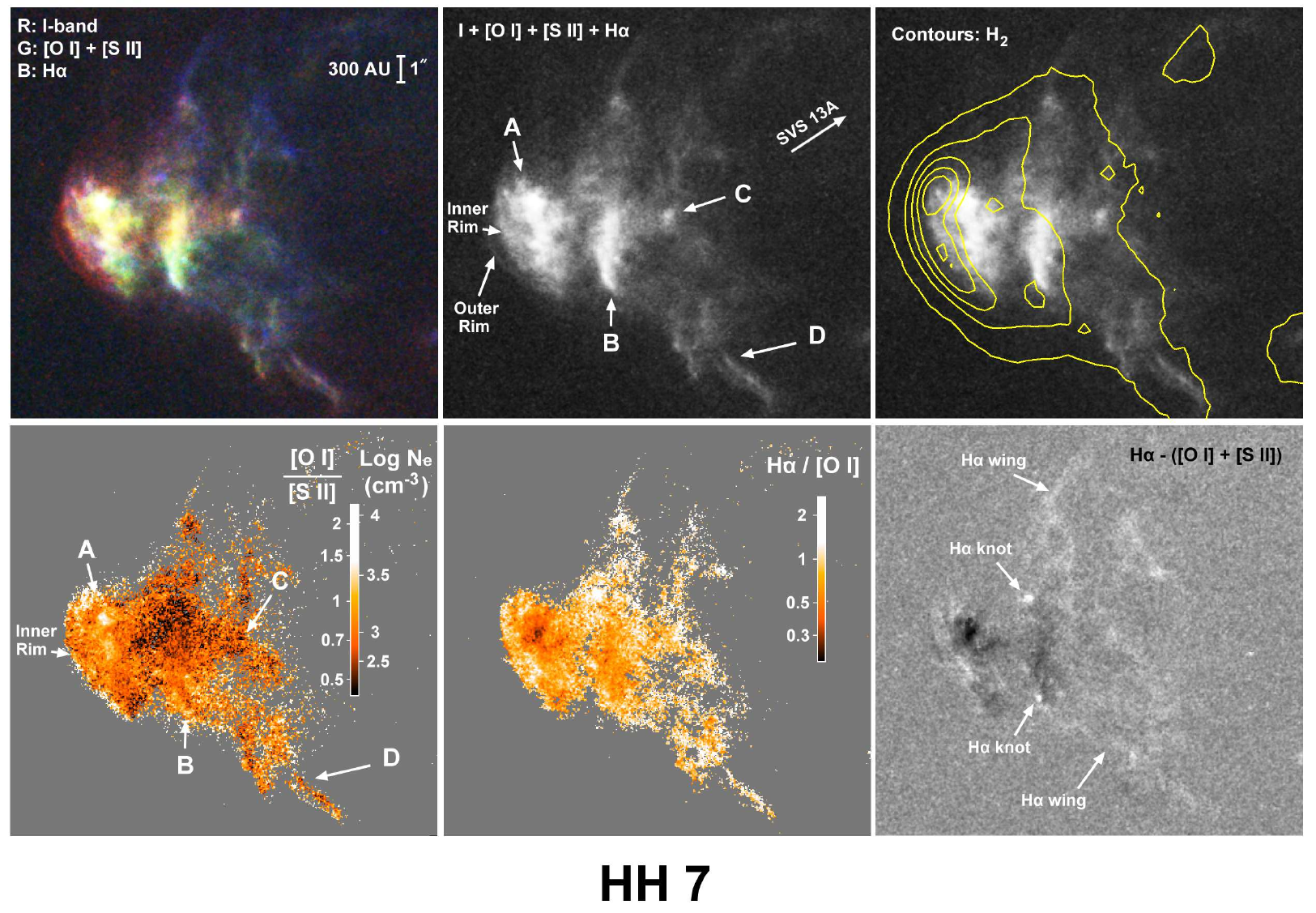}
\caption{Top-left: Color composite of HH~7, with the I-band
image (F850LP) in red, [O~I]$\lambda$6300 + [S~II]$\lambda$6716 (F631N+FQ672N)
in green, and H$\alpha$ (F656N) in blue. North is up and east to the left.
The scale bar adopts a distance of 300~pc. Top-middle: Combined I + [O~I]6300 + [S~II]6716 + H$\alpha$
image with the labeled features discussed further in the text. The direction to the exciting source
SVS 13A is shown. Top-right: Contoured
ground-based H$_2$ image superposed upon the HST optical emission.
Bottom-left: The [O~I] $\lambda$6300 / [S~II] $\lambda$6716 ratio images,
and their translations to log N$_e$ using the 8000~K curve in Fig.\ref{fig:theory}.
Grey areas have low emission line fluxes in one or both lines. Bottom-middle: A similar
ratio map for the H$\alpha$ / [O~I] $\lambda$6300 ratio.  Bottom-right: A difference image between
H$\alpha$ (white) and the cooling zone forbidden line emission ([O~I] + [S~II], black).
}
\label{fig:hh7c}
\end{figure}

\begin{figure}
\centering
\includegraphics[angle=0,scale=1.00,width=\textwidth]{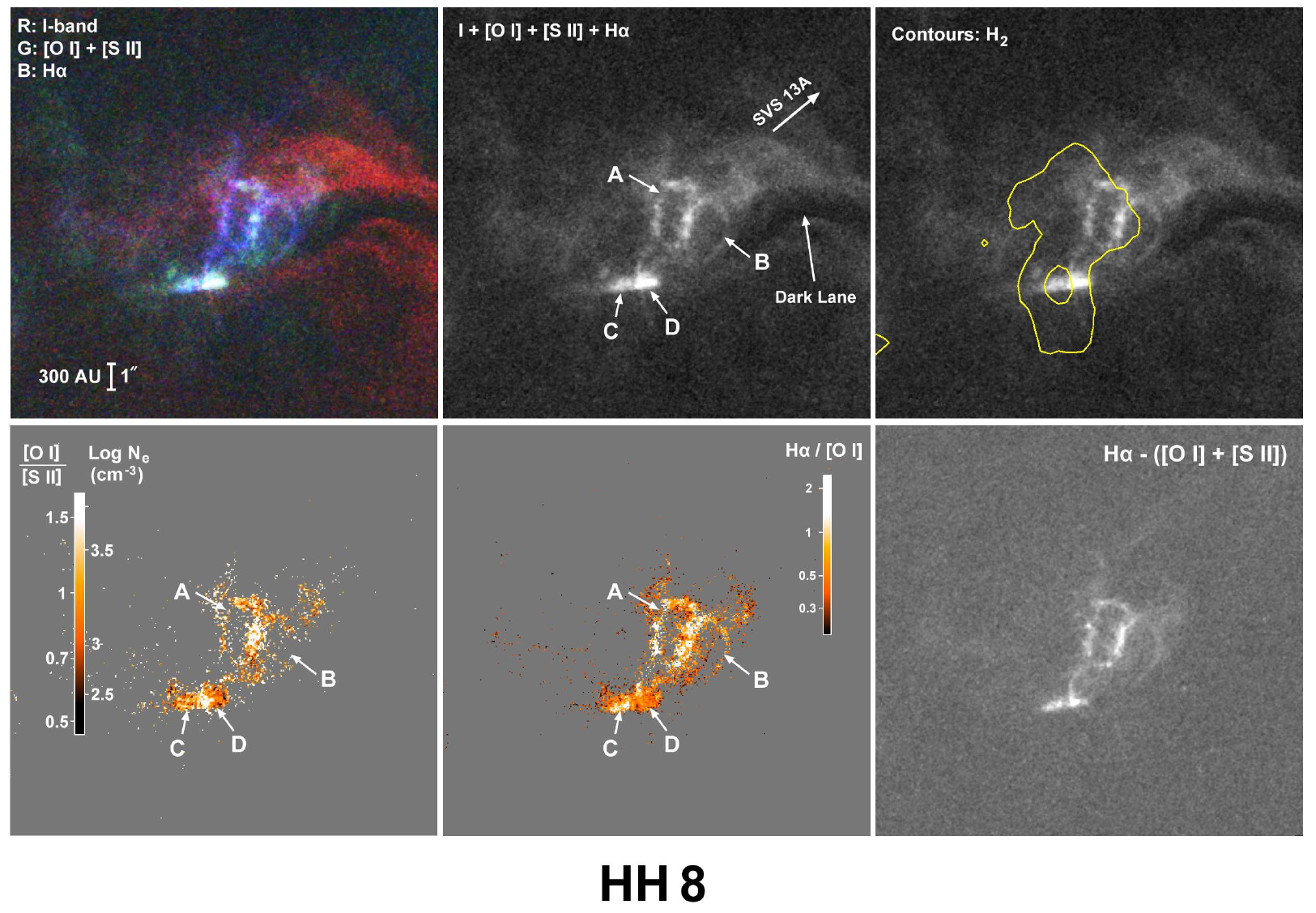}
\caption{Same as Fig.~\ref{fig:hh7c} for HH~8.
}
\label{fig:hh8c}
\end{figure}

\begin{figure}
\centering
\includegraphics[angle=0,scale=1.00,width=\textwidth]{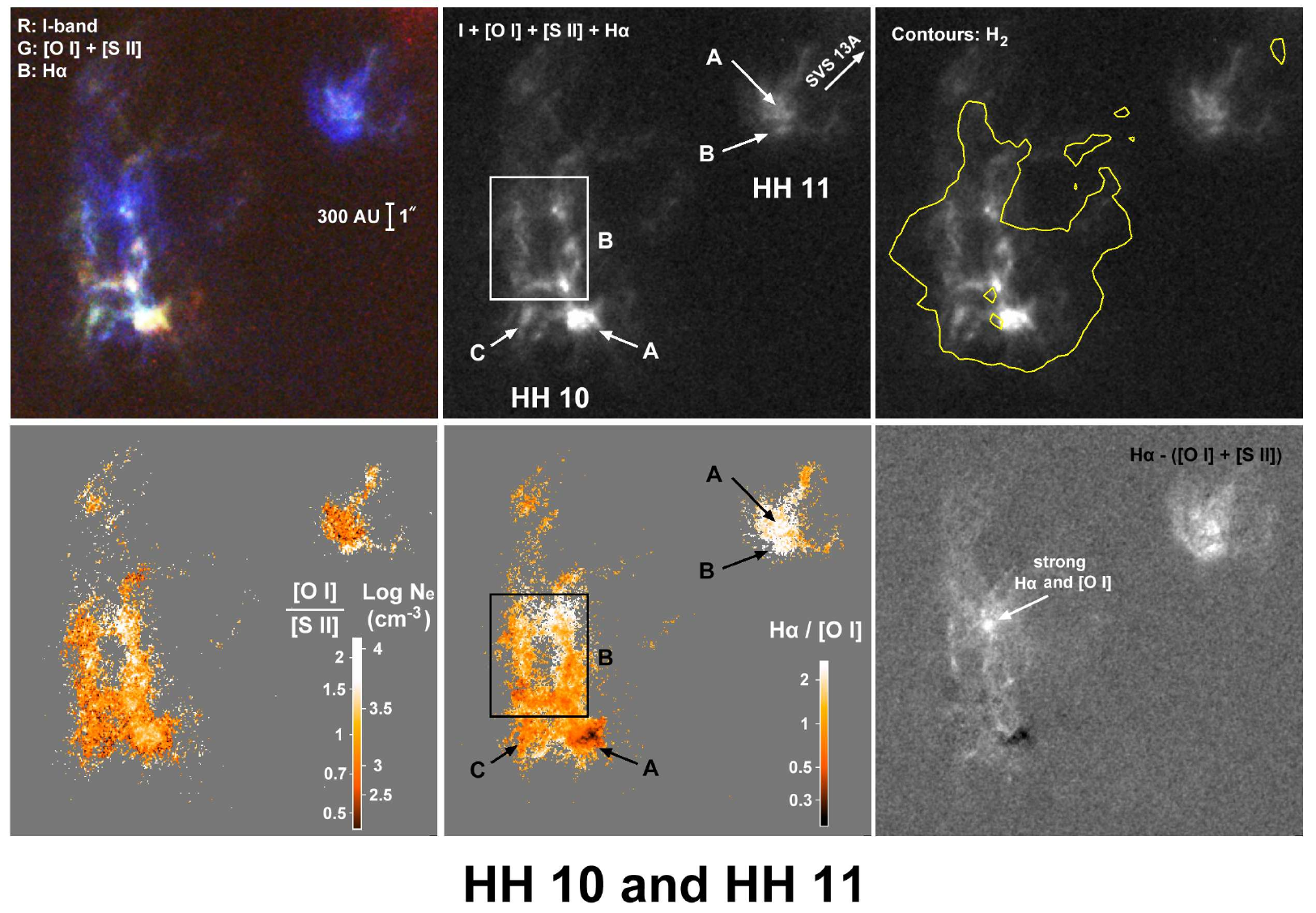}
\caption{Same as Fig.~\ref{fig:hh7c} for HH~10 and HH~11.
}
\label{fig:hh10c}
\end{figure}

The top-left panels of
Figures \ref{fig:hh7c}, \ref{fig:hh8c}, and \ref{fig:hh10c}
present color composites of the new HST images of HH~7, HH~8 and HH~10, and HH~11,
respectively.  In these composites, the I-band image (F814W; red) shows primarily
continuum sources such as stars and reflection nebulae, although HH objects also 
radiate several weak emission lines (e.g. [Ca II] $\lambda$7291, [O II] $\lambda\lambda$7321+7330,
[Ni II] $\lambda$7380, [Fe~II] $\lambda$8617, etc.) within this bandpass. The H$\alpha$ (F656N; FWHM $\sim$ 18\AA ),
[O~I] $\lambda$6300 (F631N; FWHM $\sim$ 61\AA ) and [S~II] $\lambda$6716 (FQ672N; FWHM $\sim$ 19\AA )
filters isolate single emission lines. Only relatively bright stars have enough continuum to be detected
in these filters. None of the HH objects in the field have a detectable continuum component
in the F645N filter (FWHM $\sim$ 85\AA ).
Continuum is sometimes present in the I-band images of the HH objects because
the I-band is less affected by reddening than the R-band, and because the bandpass
of the F814W filter ($\sim$ 2000\AA ) is much wider than those of the narrowband filters. 

Because the HH objects in HH~7-11 are pure emission-line sources, 
we can create difference images and emission ratio maps directly from the zero-point-corrected
H$\alpha$, [O~I] $\lambda$6300 and [S~II] $\lambda$6716 images. As noted in
Sec.~\ref{sec:intro}, [O~I] $\lambda$6300
and [S~II] $\lambda$6716 trace similar regions in the cooling zones of shocks, 
while the surface brightness of H$\alpha$ peaks in a narrow shell immediately
behind the shock for low-velocity shocks like those in HH~7-11 \citep[e.g.][]{hw15}. Hence, 
the difference image H$\alpha$ $-$ ([O~I] $\lambda$6300 + [S~II] $\lambda$6716) can
be useful for separating shock fronts from their cooling zones. Shocks should become
manifest in the subtraction images as filamentary H$\alpha$-bright structures.

Two line ratio maps are particularly useful as diagnostics in HH~7-11. 
The first of these, [O~I] $\lambda$6300 / [S~II] $\lambda$6716, is normally 
not a particularly good
density diagnostic because its value depends linearly upon the ionization fraction
of O in the cooling zone, a quantity that is affected both by the preshock
ionization fraction and by the shock velocity.
However, the shocked regions in the HH 7 - 11 outflow are special cases in that they have unusually
low-excitation optical spectra, such as [N~II] $\lambda$6583 / [N~I] $\lambda$5198+5200 $\lesssim$ 0.1,
[O~I] $\lambda$6300 / H$\alpha$ $>$ 1 and [S~II] $\lambda$6716 / H$\alpha$ $>$ 1, weak [O~II],
and no [O~III] lines \citep{bohm83,sb90}. These ratios all imply very
low H-ionization fractions of at most a few percent. 
The ionization fraction of oxygen is tied to that of hydrogen through
strong charge-exchange reactions \citep{williams73}, so the ionization fraction of oxygen is also
at most a few percent in low-excitation HH objects. Ionization effects are
unimportant for S; the low ionization threshold
of S implies it should be at least singly-ionized throughout the cooling zone, and 
the lack of any [S~III] lines implies that all S is S~II to within a few percent.

Under these mostly neutral conditions, [O~I] $\lambda$6300 / [S~II] $\lambda$6716 
becomes an excellent diagnostic of relative density throughout the flow. In fact,
the [O~I] / [S~II] ratio is superior to the standard [S~II] $\lambda$6716 / [S~II] $\lambda$6731
ratio in that it can measure electron densities two orders of magnitude above the high density
limit ($\sim$ $2\times 10^4$ cm$^{-3}$) of the [S~II] ratio.
The [O~I] $\lambda$6300 / [S~II] $\lambda$6716 ratio increases monotonically with electron density 
because the critical density of [O~I] $\lambda$6300 is much higher than that
of [S~II] $\lambda$6716.  The temperature dependence in the [O~I] $\lambda$6300 / [S~II] $\lambda$6716
line ratio is weak, and a recent large grid of shock models shows that S~II and
O~I emit at similar temperatures in the postshock region \citep[$\sim$ 8000~K; ][]{hw15}. 

We must scale [O~I] $\lambda$6300 / [S~II] $\lambda$6716 by the 
abundance ratio of O and S to use it to calculate electron
densities quantitatively, though the relative densities across a given 
map are unaffected by this normalization constant. In the HH 7-11 region we can determine
the O / S abundance ratio by combining our data with published ground-based spectra.
The observed [S~II]$\lambda$6716 / [S~II]$\lambda$6731 ratio in a 6.9 arcsecond
aperture centered on HH~11 is 0.82 \citep{bohm83,sb90}, implying 
log(N$_e$) = 3.06 using modern values of collision strengths \citep{sd94}
for T $\sim$ 8000~K.  For temperatures between 7000~K and 9000~K
characteristic of shocked cooling zones, 
ionization fractions between 0 and 0.1, and an O / S gas abundance
ratio of 35 (log O = 8.87, log S = 7.33  with log H = 12) the 
[O~I] $\lambda$6300 / [S~II] $\lambda$6716 ratio in a 5-level atom calculation
for N$_e$ = $10^3$ cm$^{-3}$ gives 0.71, with a scatter of $\pm$ 15\%.

The observed ratio of [O~I] $\lambda$6300 / [S~II] $\lambda$6716 in our HST images
over the same region is 0.58 $\pm$ 0.08.
Dereddening this ratio with E(B$-$V) = 0.62
\citep[][corresponding to C$_{H\beta}$ = 0.98]{bohm83},
we get [O~I] $\lambda$6300 / [S~II] $\lambda$6716 = 0.65 $\pm$ 0.08 for HH~11, in
agreement with the value of 0.71 expected from the observed
electron densities.  The same procedure using the ground-based spectra of HH~7 \citep{sb90}
gives similar agreement.  Hence, our observations are consistent with a
normal abundance ratio for O / S in the HH 7-11 region.

\begin{figure}
\centering
\includegraphics[angle=0,scale=1.00,width=\textwidth]{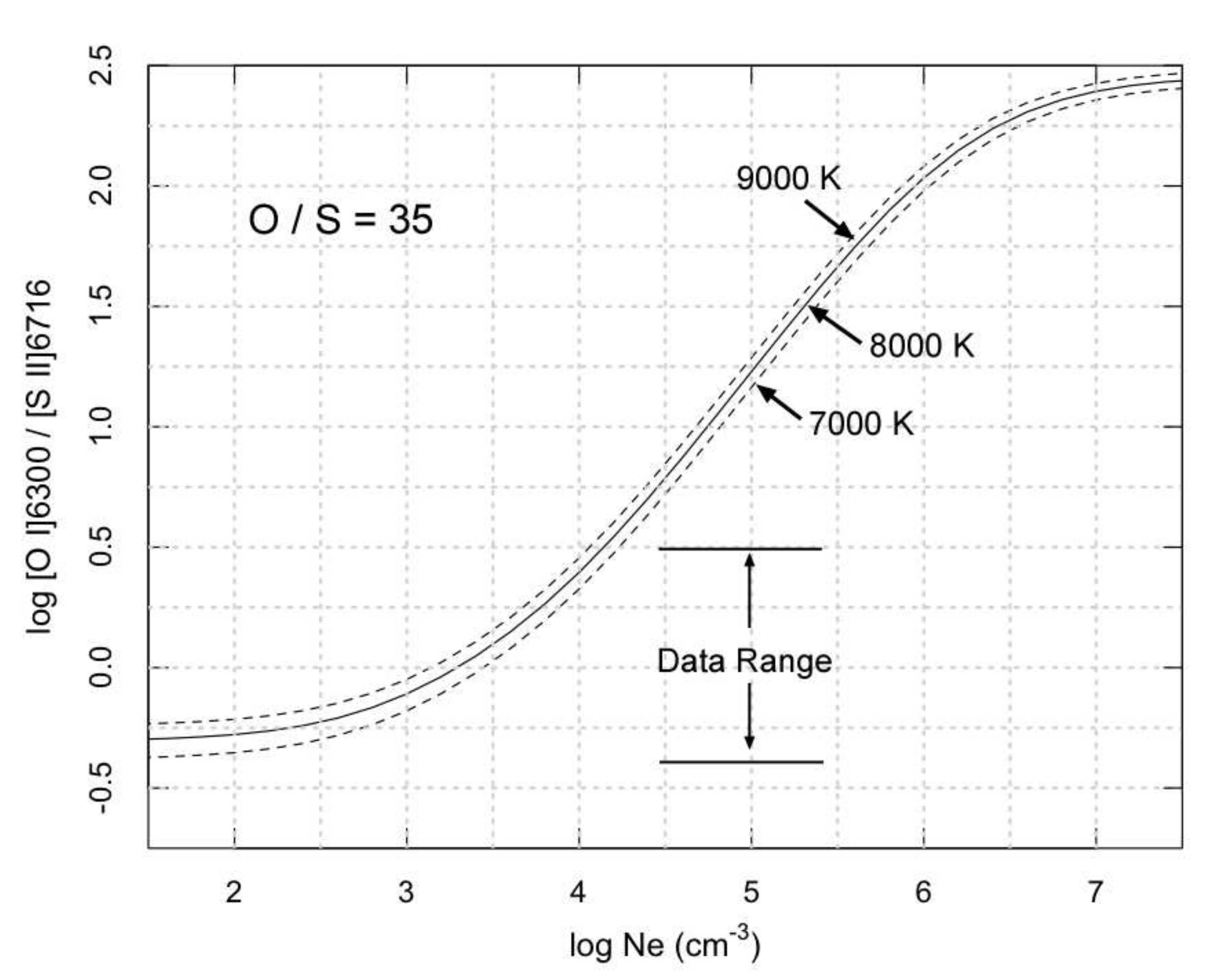}
\caption{Diagnostic diagram for the [O I] $\lambda$6300 / [S II] $\lambda$6716 line
ratio for three temperatures and an abundance ratio O / S = 35.
The curves rise monotonically with
density. The range of the ratio measured in the images is shown.
}
\label{fig:theory}
\end{figure}

Diagnostic curves for the [O~I] $\lambda$6300 / [S~II] $\lambda$6716 emission line
ratio are shown in Fig.~\ref{fig:theory} for three temperatures,
assuming neutral O, singly-ionized S, and an O / S abundance ratio of 35. 
As a check, the observed ratios throughout the image fall within
the allowed range for these abundances,
and indicate electron densities that range from the low density
limit ($\lesssim$ 100 cm$^{-3}$) to $\sim$ $2\times 10^4$ cm$^{-3}$
(Fig.~\ref{fig:theory}).

The H$\alpha$ / [O~I] $\lambda$6300 ratio is another important line ratio
we can measure directly from our images.
\citet{bohm83} noted that both HH~7 and HH~11 had unusually low-excitation
spectra for HH~objects, with H$\alpha$ / [O~I]$\lambda$6300
$\lesssim$ 1.  Integrated over the entire object, a
lower value for H$\alpha$ / [O~I]$\lambda$6300 implies a lower
shock velocity \citep[Fig.~8 of][]{hrm94}, and this 
constraint together with the other observed 
line ratios implies shock velocities V$_S$ $\sim$ 30 km$\,$s$^{-1}$ according to the 
diagnostic curves of \cite[][cf. their Fig.~5]{ds17}. 
For images like ours where the cooling zones are resolved
spatially, a lower value of H$\alpha$ / [O~I]$\lambda$6300
signifies the location of the T $\sim$ 8000~K gas.

Background-corrected ratio images for [O~I] $\lambda$6300 / [S~II] $\lambda$6716 
and for [O~I] / H$\alpha$ appear in Figs.~\ref{fig:hh7c}, \ref{fig:hh8c}, and
\ref{fig:hh10c} along with multicolor composites that employ the same
color schemes as in Fig.~\ref{fig:overview}.  The ratio images exclude points,
shown as grey, where the count rate
is lower than a threshold in either the numerator or the denominator. 
We chose threshholds of 0.0035, 0.0030 and 0.0030 electrons/second
for [O~I], [S~II], and H$\alpha$, respectively, because they produced
ratio images with minimal scatter while still spanning 
each of the HH objects in the field.

\subsection{HH 7}

Fig~\ref{fig:hh7c} shows that the three brightest components of
HH~7, labeled A, B, and C in the Figure, each have an overall bow-shape
characteristic of a flow moving from WNW to ESE. A narrow outer rim
visible only in the I-band precedes knot A, while a similar inner rim 
defines the eastern edge of the H$\alpha$ and forbidden line emission in this object.
None of the other HH objects show similar rims in the I-band.
The I-band rim has no counterpart in the images of
the bright red emission lines of [O~I], [S~II] and H$\alpha$. Hence, this feature
must emit mostly continuum, and in star formation regions this typically 
occurs along the walls of a cavity that reflects the light of
a nearby protostar. 
The H$_2$ emission here is also unique for the region, appearing as
a sharp arc that follows along the I-band rims. Unfortunately, the resolution of
the ground-based H$_2$ is too low to determine any spatial offsets between
the H$_2$ and I-band rims.  The I-band and H$_2$ rims likely mark the
terminus of the flow, with the I-band rim arising mostly from light reflected
from shells of dust at the end of the cavity, while
the H$_2$ could come from a precursor to the optical shock in knot A.
This area has great potential for elucidating the physics of
molecular shock waves, and should be a prime target for future study with JWST.

The brightest portion of HH~7A is located about one arcsecond to its 
west of the inner rim, and has a complex, serrated morphology anchored
by a bright knot at its northern end (Fig.~\ref{fig:working-surface}).
A complex region of emission situated
inside of a bow shock like this is expected of a working surface, where
jet material decelerates.  The brightest knot splits into three
components upon close inspection and appears to have a wake to the southwest,
mainly visible in H$\alpha$. The knot stands out as the region of
highest electron density ($\gtrsim 10^4$ cm$^{-3}$) in the 
[O~I] / [S~II] ratio map, and also has a relatively low
H$\alpha$ / [O~I] ratio.  The knot also radiates in the I-band, 
but as this part of the spectrum also emits forbidden lines
we cannot tell what fraction of the I-band emission in this object arises
from lines and what fraction comes from continuum.

Like HH~7A, HH~7B also has a bow-shape, and it is possible to trace the
wings of the bow shock in H$\alpha$ $\sim$ 5 arcseconds to the north 
of the object, as well as $\sim$ 5 arcseconds to the southwest, where
it overlaps with HH~7D at the edge of the flow. These H$\alpha$ wings
are easiest to follow in the difference image in the lower right
panel of Fig.~\ref{fig:hh7c}. HH~7B has two bright H$\alpha$ knots of diameter $\sim$ 0.3
arcseconds that show up well in the difference image. The inner bow HH~7C
is less-distinct, but has the same overall appearance as HH~7B. 

\subsection{HH 8}

The emission line images of H$\alpha$, [S~II] and [O~I] reveal 
HH~8 to be a remarkable object that consists of two distinct loops,
(denoted A and B), situated above a bright linear feature (knots C and D;
Fig~\ref{fig:hh8c}). The periphery of knot A breaks up into multiple knots,
the brightest and densest of which occurs where loop A intersects with loop B.
On small scales, knot C breaks into two components and knot D into three, the 
easternmost of the three showing distinctly higher [O~I] / [S~II] than
the other two.  The H$_2$ contours peak between knots C and D.
The forbidden-line emission is much weaker compared with H$\alpha$
in HH~8 than it is in HH~7, so the difference image of HH~8 is not particularly illustrative.
The I-band image shows what appears to be an extended background nebula,
probably the rear wall of a cavity evacuated by the jet. A distinct dark
lane cuts across this nebula, but does not seem to affect HH~8B, as expected
if the HH object lies above the obscuring lane and the background nebula.

\subsection{HH 9}

HH~9 is a rather indistinct
source located along the northeastern edge of the cavity defined by the
reflected light in the I-band image (red in Fig.~\ref{fig:overview}).
It has a diameter of just over an arcsecond, and has little substructure.
The HH object is visible in H$\alpha$ and [S~II] $\lambda$6716, and
also detected faintly in [O~I] $\lambda$6300. 

\subsection{HH 10}
 
Like HH~8, HH~10 has a remarkable, nearly rectangular loop (HH~10B)
located north of bright knots of emission (HH~10A and HH~10C; Fig.~\ref{fig:hh10c}). 
Like HH~8A, the HH~10B loop breaks up into multiple knots, including one
knot that emits strongly in H$\alpha$ and in [O~I] relative to [S~II]
(a blue-green dot in the upper left panel and white dot in the lower right panel).
The densities are on average higher along the west side of HH~10
and lower along the east side. Knot A is notable in that it has
the strongest forbidden lines relative to H$\alpha$. Diffuse H$_2$
emission spans the extent of HH~10.

\subsection{HH 11}

HH~11 is the highest-excitation object in the chain, with the strongest
H$\alpha$ emission relative to [O~I] and [S~II].   The H$\alpha$ separates into
two distinct objects, HH~11A, a tadpole-shaped feature located closer to
the driving source, and HH~11B, a shell-like object that precedes
HH~11A in the flow. Forbidden lines are essentially
absent in HH~11B, and only weakly present in HH~11A. There is no
H$_2$ emission, and the I-band shows only faint features as
one would expect from weak emission lines in the band. 
Both HH~11 and HH~10 are located in a hole in the
reflected-light cavity seen in the I-band (Fig.~\ref{fig:overview}).

\section{Distances and Resolved Young Stellar Objects in the Field}
\label{sec:ysos}

The NGC~1333 region contains 205 candidate young stellar objects according to a
recent survey \citep[][see also \citet{rebull15}]{foster15}, and 14 of these are visible in our images.
The candidate young stars are all 2MASS infrared sources,
and most have designations from previous near-IR catalogs \citep[e.g.][sources
designated as ASR]{aspin94}. There are several molecular outflows
and H$_2$ jets within NGC~1333 \citep{plunkett13,garden90}.

Figure~\ref{fig:gaia} displays a histogram of the distances determined
by GAIA-2 catalog for members of NGC~1333 \citep{gaia1,gaia2}.
Of the 205 sources in \citet{foster15}, 114 were detected by GAIA, 99 had
parallax measurements, and 50 of these had parallax errors $<$ 10\% . Using
10000 bootstrap samples, the median value for the distance is 296 $\pm$ 5 pc.
This distance is consistent with previous estimates based on Hipparchos data
\citep{deZ99,belikov02}. In this paper we adopt 300~pc as the
distance to the HH~7-11 outflow.

A large-scale imaging survey of this region with HST
in the near-IR taken nearly simultaneously with our observations
(GO 15153; D.~Watson PI) will investigate the stellar objects in more detail.
However, a few of the young stars have resolved circumstellar nebulae in
our images that warrant discussion here. 
SVS~13A shows the most interesting morphology of the young stars in our images (Fig.~\ref{fig:stars}).
The I-band reveals at least three arc-shaped nebulosities, leaving
the impression of two elliptical-shaped cavities that originate from SVS~13A. The
smaller cavity has PA = 162 $\pm$ 2 degrees and major axis 1.9 $\pm$ 0.1 arcseconds, and the
larger one PA = 156 $\pm$ 2 degrees with major axis 2.8 $\pm$ 0.1 arcseconds. The
sizes of the minor axes are 1.0 and 1.6 arcseconds, respectively. These cavities
correspond with those observed in H$_2$ by \cite{hodapp14}, who also observed a third,
smaller cavity close to the source. These observations imply
a scenario where multiple ejections driven from
SVS~13A in a southeasterly direction leave behind fossil cavities.

The [O~I] $\lambda$6300 and [S~II] $\lambda$6716 images show that a microjet extends $\sim$ 0.3 arcseconds
to the southeast of SVS~13A at PA = 145 degrees, in the general direction of the HH~7-11 outflow.
This jet is also seen in Fe~II 1.64$\mu$m AO datacubes taken with Keck \citep{hodapp14},
where the radial velocity of $\sim$ $-140$ km$\,$s$^{-1}$ shows some evidence for
variability. SVS~13A itself is clearly variable between the two HST epochs,
having faded by 0.37 $\pm$ 0.1 magnitudes
between 1998 and 2018 in the H$\alpha$ images when compared with the two nearby
bright stars ASA~2 and ASA~3 (all unsaturated in the narrowband images). The
decline in brightness is similar, 0.39 mag, in the narrowband O~I filter.

Our HST images also show extended arcuate nebulosities
typical of evacuated cavities and embedded objects
around two stars to the northeast of SVS~13A (Figs.~\ref{fig:overview}, \ref{fig:stars}). 
The I-band images of the first of these, ASR~7, shows a broad arc 
centered on the source and extending several arcseconds to the west. 
The nebula is significantly brighter on its north side than its south side.
ASR 7 (2MASS J03290575+3116396) is an X-ray source and
has very red colors in the
near-IR \citep[K $\sim$ 10, J$-$K $\sim$ 4.5; ][]{winston10}.
Spectra reveal a rapidly rotating early M/late-K star, with a significant
amount of veiling at H \citep{foster15}. 
Two fainter stars noted by \citet{oasa08} are visible in our images as well,
situated about 2 arcseconds to the NE and to the SE of ASR 7.
The other star in this region with a circumstellar nebula in our images is ASR~105.
This nebula has a compact C-shape that extends $\sim$ 1 arcsecond
to the northeast of the source. ASR~105 lies near the
brown dwarf limit, with a spectral type of M6 and a signficant veiling
in the near-IR indicative of ongoing accretion \citep{greissl07}.

Finally, ASR~2 (2MASS J03290289+3116010), located $\sim$ 11.5 arcseconds to the WSW of SVS~13A,
resolves cleanly into a close binary in our HST images in Fig.~\ref{fig:stars}. 
Using DAOPHOT, we find the secondary is positioned nearly due
north of the primary at PA=0.8 $\pm$ 2.1 degrees, separation = 0.15 $\pm$ 0.01 arcseconds,
with a magnitude difference of 0.98 $\pm$ 0.07 relative to the primary in the R-band continuum image.
We used the other nearby bright star, ASR~3 (2MASS J03290216+3116114),
as the point-spread-function star for the photometric and positional measurements.
The secondary of ASR~2 is distinct in all the different filters, though is more difficult to
see in the I-band image because the primary is saturated in those exposures.
\cite{foster15} characterized the combined spectrum of both components of ASR~2
as a slowly-rotating, weakly-veiled T-Tauri star with an effective temperature of 3760~K
and a radial velocity consistent with membership in NGC~1333.
We checked to see if any orbital motion might have occurred between our
observations and those of the previous HST images taken in 1995 and 1998,
but the binary nature of ASR~3 was unclear in the archival
WFPC2 images owing to the larger pixel sizes in those data. This object would be
a good one to monitor in the future for orbital motion as a means to test
the pre-main-sequence evolutionary tracks for very low mass stars.

%
%
%
%
%

\section{Proper Motions and Variability of the HH Objects}
\label{sec:motions}

The HH~7-11 outflow has been targeted for narrowband optical imaging with WFPC2 on
HST in two previous campaigns. In 1995, program GTO-5779 observed the region 
in H$\alpha$ (F656N), [S~II] (F672N), and I-band (F850LP), but a positioning error
left all of HH~7 outside the field of view. Just over three years later, GO-6868
observed the same region in H$\alpha$ (F656N), [S~II] (F672N), and [O~I] (F631N),
but these observations again missed the tip of the bow shock in HH~7. However,
both sets of observations imaged HH~8, HH~9, HH~10, and HH~11, so we can
combine the previous images with ours to measure high-precision
proper motions for these objects.  It is possible to measure proper motions for
some interior portions of the HH~7 bow shock from GO-6868, but not for the apex.  No proper motions
are evident in the three year time-interval between the GTO-5779 and GO-6868 images of
HH~8, HH~9, and HH~10. These images are similar enough that we
combined them into a single epoch to increase the signal-to-noise.
A proper motion of a few pixels directed away from SVS~13 is present 
between GTO-5779 and GO-6868 in HH~11.

Images in GTO-5779 and GO-6868 (both referred to here as epoch 1) were
not dithered when acquired, and so have somewhat poorer
spatial resolution than we achieve in our current (epoch 2) dataset. This effect slightly blurs
the epoch 1 data relative to epoch 2 but does not affect the proper motion measurements.
Alignment between the three datasets is a more difficult issue. The bright stars SVS~13, ASR~2, and
ASR~3 are present in all the images, but the field does not have any visible stars to
the southeast in the vicinity of HH~7.  Moreover, SVS~13 has a variable microjet 
in H$\alpha$ and [S~II] that affects its photocenter (Fig.~\ref{fig:stars}), while
ASR~2 is a binary (Sec.~\ref{sec:ysos}; Fig.~\ref{fig:stars}) and is saturated in most exposures. 
A few stars are available for alignment to the north and west of SVS~13, but these are
out of the field of view for some of the observations, and faint in the narrowband filters.
For these reasons we used the orientation and pixel scale recorded when the observations
were taken to rotate and scale the images, and relied upon the HST reduction pipeline
to correct for spatial distortion. After rotation and scaling, the zero-points of the world-coordinate
systems aligned to within an arcsecond or so, and we did a final spatial shift to align the
photocenter of ASR~3 in each image. The resulting alignments should be accurate to
within 1 WFPC2 pixel (0.1 arcseconds), and may be better than that, but we cannot
confirm this owing to the lack of alignment stars.  

The epoch 1 and epoch 2 images in H$\alpha$ are presented side-by-side in Fig.~\ref{fig:time}. 
We measured proper motions for the regions indicated in each of the light-blue boxes.
Each figure also shows several white boxes, which we present as fiducials that aid
in viewing the motions between epochs. If the reader expands the images to a
comfortable size and then looks quickly between the left and right panels, it should
become clear both how the emission translates in bulk relative to the fixed 
white and blue boxes, and the degree to which the
objects vary morphologically in the 20-year timespan between the epochs.
The reader may also refer to the on-line animated figure, which switches back and forth
between the epochs for HH~7 and HH~8 (Fig.~\ref{fig:movie7}) and for
HH~10 and HH~11 (Fig.~\ref{fig:movie11}).

\subsection{HH 7}

Our proper motion measurements in HH~7 are limited because both epoch 1 images
omit its most recognizable feature, the leading bow shock. Nevertheless,
a few filamentary structures that follow in the wake of the main bow shock retain
enough of a coherent morphology to be useful for proper motion measurements. The four
areas shown in Fig.~\ref{fig:time} have small, but nonzero proper motions
to the east, as suggested by the orientation of the leading bow shock. The fastest object
is the arc-shaped bow HH~7C, at 38 km$\,$s$^{-1}$, with the slowest
being HH~7E, at 12 km$\,$s$^{-1}$. Small proper motions
in declination are near the limit of confusion with morphological changes
in the shocked gas.

\subsection{HH 8}

Emission within the blue rectangle that defines the extent of HH~8 in Fig.~\ref{fig:time} 
exhibits no detectable proper motion ($\lesssim$ 6 km$\,$s$^{-1}$, Table~1). However, HH~8 shows remarkable
structural changes between the two epochs. Both loops HH~8A and HH~8B expand
noticeably in the 20-year interval, HH~8A by 25 $\pm$ 5 km$\,$s$^{-1}$
along its northern and southern portions, and HH~8B by about half this speed
along its western side. The interface between the two loops is stationary. 
The bright knots HH~8C and HH~8D located to the southeast of the two loops have no
bulk proper motions, but their relative brightnesses have changed, with knot HH~8C brighter than
knot HH~8D in the late 1990's and knot HH~8D the brighter of the two now. HH~8D is
notable for its current strong forbidden line emission and high electron density,
particularly along its eastern edge (Fig.~\ref{fig:hh8c}). The sharp H$\alpha$
filament to the north of HH~8A is now more diffuse.

\subsection{HH 9}

Figure \ref{fig:overview} shows HH~9 to be an indistinct source situated along
the edge of a cavity outlined by the reflected light present in the I-band. 
The source is a diffuse structure that shows no significant proper motions
or morphological changes.

\subsection{HH 10}

Like HH~8, HH~10 has no bulk proper motion ($\lesssim$ 6 km$\,$s$^{-1}$), but exhibits structural
variability. The large loop HH~10B retains its overall shape between the epochs,
but several small clumps have now appeared on both the east and west sides of the loop.
One compact knot has largely replaced the three elongated ones that
defined the northwestern corner in the epoch 1 image. At the base of the loop, the
three knots that make up HH~10A have been replaced by a single knot. One gets a
general impression that the large loop HH~10B has become narrower 
shifted a bit along the direction of the flow. This impression is driven largely
by the appearance of the western side of HH~10, which curves to the west in epoch
1 but becomes more straight in epoch 2.

\subsection{HH 11}

HH~11 has by far the highest proper motions in the region, $\sim$ 1.2 arcseconds
over the $\sim$ 20 year time interval. Motion is directed away from SVS~13.
Morphological changes in HH~11 are rather minor.
The leading shell of H$\alpha$ separates from the trailing blob of
H$\alpha$ and forbidden line emission by $\sim$ 0.3$^{\prime\prime}$ in the new
images, somewhat more than in the epoch 1 data where there is no clear space
between these components. But otherwise, HH~11 retains its shape well over the 
period.  

However, HH~11 is noticably fainter in the most recent epoch relative to the other HH objects.
All the HH objects appear brighter by $\sim$ 30\%\ {\it relative to the stars} in
the second epoch H$\alpha$ images, but this increase simply results from the narrower bandpass
of the F656N filter in the WFC3 images relative to the WFPC2 images (the effect is
reversed in F631N, where WFC3 has the broader bandpass). However,
flux differences between HH objects {\it are} significant because even the narrowest
WFC3 bandpasses include all the emission from HH~11, the most blueshifted object.
For example, the transmission curve for F656N in WFC3 drops off only when
velocities are more blueshifted than
$\sim$ $-$450 km$\,$s$^{-1}$, a factor of two higher than the most blueshifted
emission from HH~11. We find that the H$\alpha$ flux from HH~11 has declined by 36\%\
$\pm$ 10\%\ in the 20 years between the epochs.  The decline in [O~I] is even 
greater, 50\%\ $\pm$ 10\%.

\section{Discussion}
\label{sec:discussion}

\subsection{Orientations, Deflections, Mass Loss Rates, and The Rings of HH~8 and HH~10}

It has been standard practice in the field to infer the orientation angle of
jets relative to the line of sight by simply combining proper motion and
radial velocity measurements, and this procedure works well as long as the HH object
behaves like a bullet that propagates through the surrounding medium. 
However, the emission in HH~8 and HH~10 arises from loop-shaped structures
that have no substantial proper motions, yet exhibit significant radial
velocities and line widths \citep[$\sim$ 50$-$70 km$\,$s$^{-1}$; ][]{sb87}. A simple
way to explain these observations is to have the shocked gas arise along
an interface where the jet has pierced through or
has been deflected from a sheet of ambient material. In
such a scenario, jet material is continuously shocked and possibly redirected
at these interfaces. The interfaces will not exhibit any bulk motion because
they are fixed to the sheet of material, and any expansion of the
rings results from the expansion of the shock within the sheet. 

\citep{raga96} modeled the passage of a jet through a clumpy medium, and
identified HH~7-11 as a possible example of such a flow, a conclusion now
strongly supported with the new Hubble images. The numerical models show that
clumpy media produce a variety of morphologically complex time-dependent shock
waves as the jet deflects from clumps and the clump shapes become modified by
shock waves. In general, shocks in the clumps have small proper motions, as
observed. No loops like we observe in HH~8 and HH~10 are obvious in the
simulations, but these could very well appear in higher spatial resolution
models or with different sets of initial conditions for the clump densities and
geometries. Situations where jets interact with clumps should be common, as jets
often extend for parsecs \citep{reipurth97} and far-IR images reveal many
complex filamentary structures in star forming regions, including NGC~1333 \citep{herschel}.

The shocks in HH~8 and HH~10 appear to be redirecting the outflow.
Together, HH~11, HH~10, HH~8, and HH~7 outline a flow that is initially oriented to the
southeast, but points almost due east by the time the flow reaches HH~7 (Fig.~\ref{fig:cartoon}).
The southern edges of the rings HH~8D and HH~10A, have low-excitation (high [O~I] / H$\alpha$) spectra and
also emit in H$_2$, implying a dense layer along that side, exactly what
is needed to deflect the jet in the observed direction (Figs. 
\ref{fig:hh8c}, \ref{fig:hh10c}). 
The clumpy morphological changes in HH~8 and HH~10 make sense as transient
phenomena that occur as the jet deflects from the obstacles.
A reflection nebula with a remarkable curved dark lane
associated with the loops in HH~8 in the I-band image (Figs.~\ref{fig:overview},
\ref{fig:hh8c}) is further evidence for a stationary obstacle at this location.  

HH 8 and HH 10 are good examples as to why {\it using line
luminosities to infer mass-loss rates in jets can give erroneous results.}
These HH objects are stationary, and emit only because the jet shocks material
along its periphery. If the sheets weren't
there, no emission would occur and the line luminosities would be zero.
In the current configuration, jet material that radiates
in HH~10 and HH~8 will radiate again when it encounters HH~7 
further down the flow, so in this case an atom in the flow will experience multiple
heating and cooling events.  Using luminosities to measure mass loss rates
only makes sense if either (a) the jet measurements are close enough 
to the source where one might argue that the entire flow radiates
before any material gets shock-heated and entrained, or (2) the emission
only comes from bullets of ejected gas, and not from shocked ambient gas. 
However, the former case is sensitive to the beam size, as larger beams
will include more emission, and the
second scenario may not produce any observable emission,
because only velocity differences produce shocks. A bullet could propagate
along a jet and be invisible until it encounters slower material. Such a
situation has been observed in HH~1, for example \citep{hartigan11}.
The best scenario is probably one where the jet is resolved spatially,
and densities and velocities can be measured close to the source
before encounters with the surrounding medium complicate the situation.

\subsection{HH~7, HH~11, SVS~13A, and the Nature of the Outflow}

HH~7 appears to be the terminus of the outflow, as
the HST images show no optical emission beyond HH~7 that would indicate a continuation
of the flow.  This conclusion is supported by the presence of a
large H$_2$ bow shock in HH~7 (Fig.~\ref{fig:hh7c}).
There is also no optical emission from the redshifted side of the molecular flow,
though the extinction there is undoubtedly large.

\begin{figure}
\centering
\includegraphics[angle=0,scale=1.00]{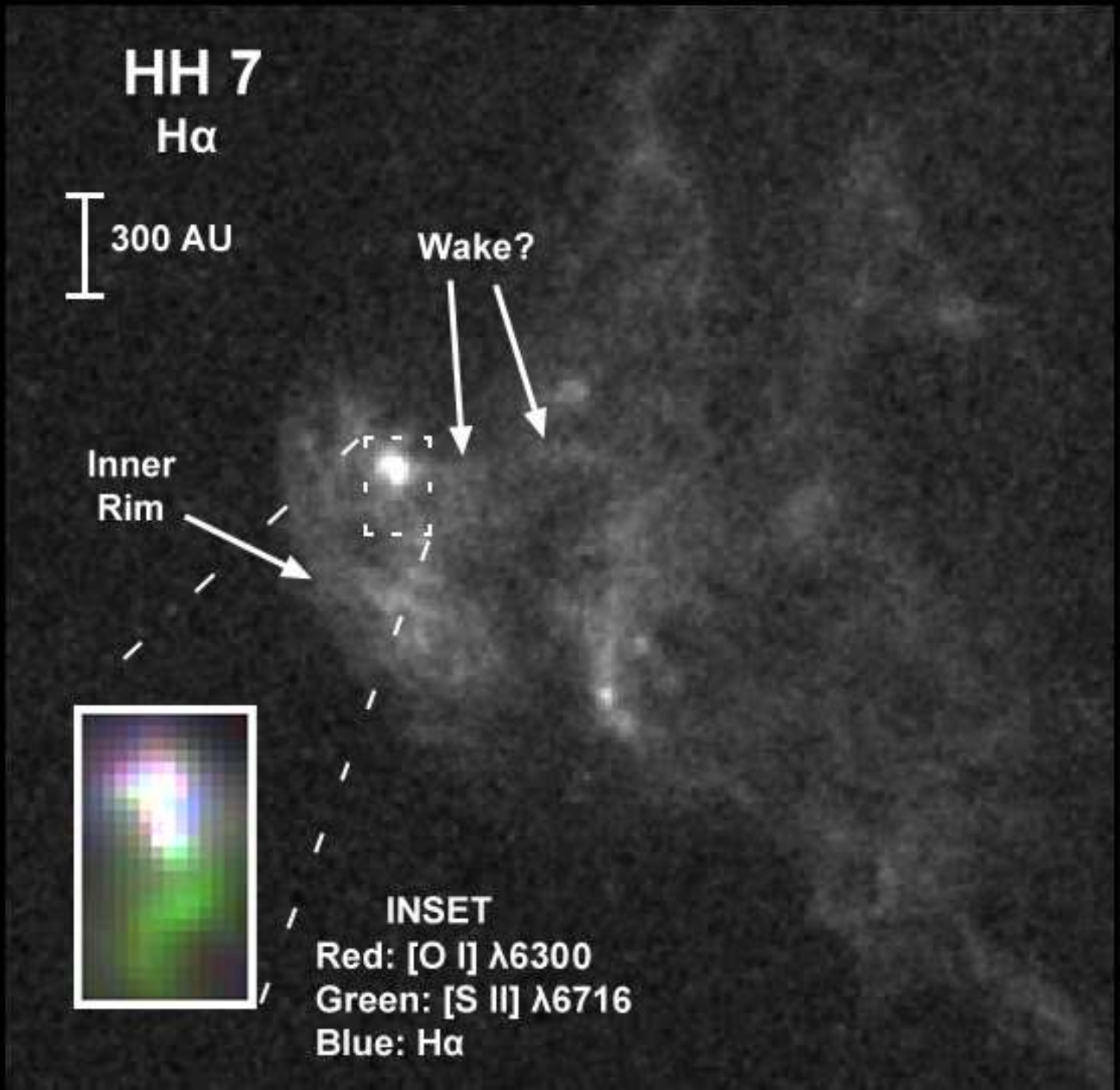}
\caption{Epoch 2 images of HH~7 (greyscale; H$\alpha$) and of the
bright knot embedded within its leading bow shock (color inset; composite). 
The bright knot shows structure down to the resolution limit of HST.
The bright knot and its possible wake suggest the jet has broken into small
clumps by the time it enters the leading HH~7 bow shock. The inner rim marks
the extent of the optical line emission (see also Fig~\ref{fig:hh7c}).
}
\label{fig:working-surface}
\end{figure}

\begin{figure}
\centering
\includegraphics[angle=0,scale=1.00,width=\textwidth]{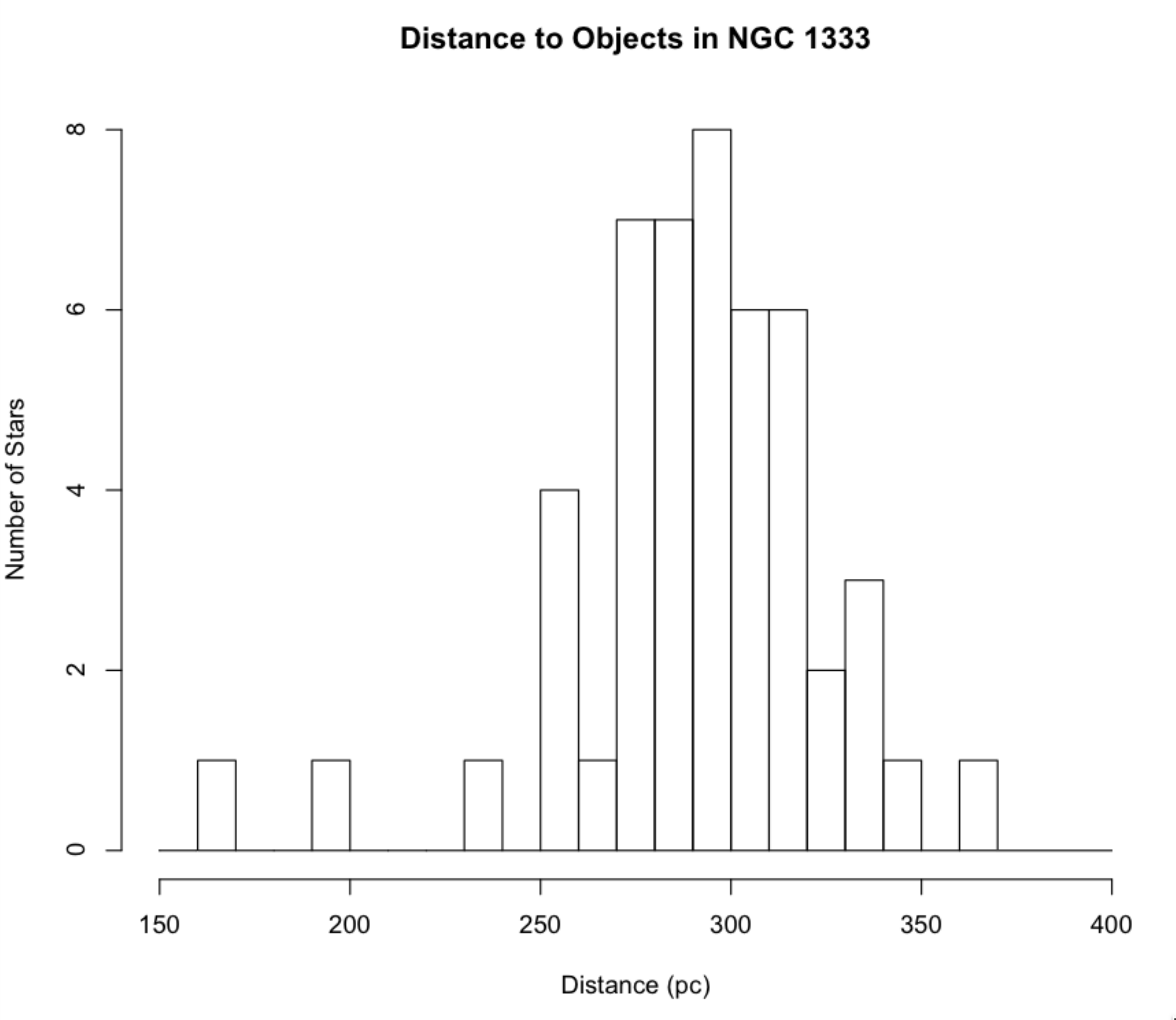}
\caption{GAIA-DR2 distances for the sources in \citep{foster15}.
}
\label{fig:gaia}
\end{figure}

\begin{figure}
\centering
\includegraphics[angle=0,scale=1.00,width=\textwidth]{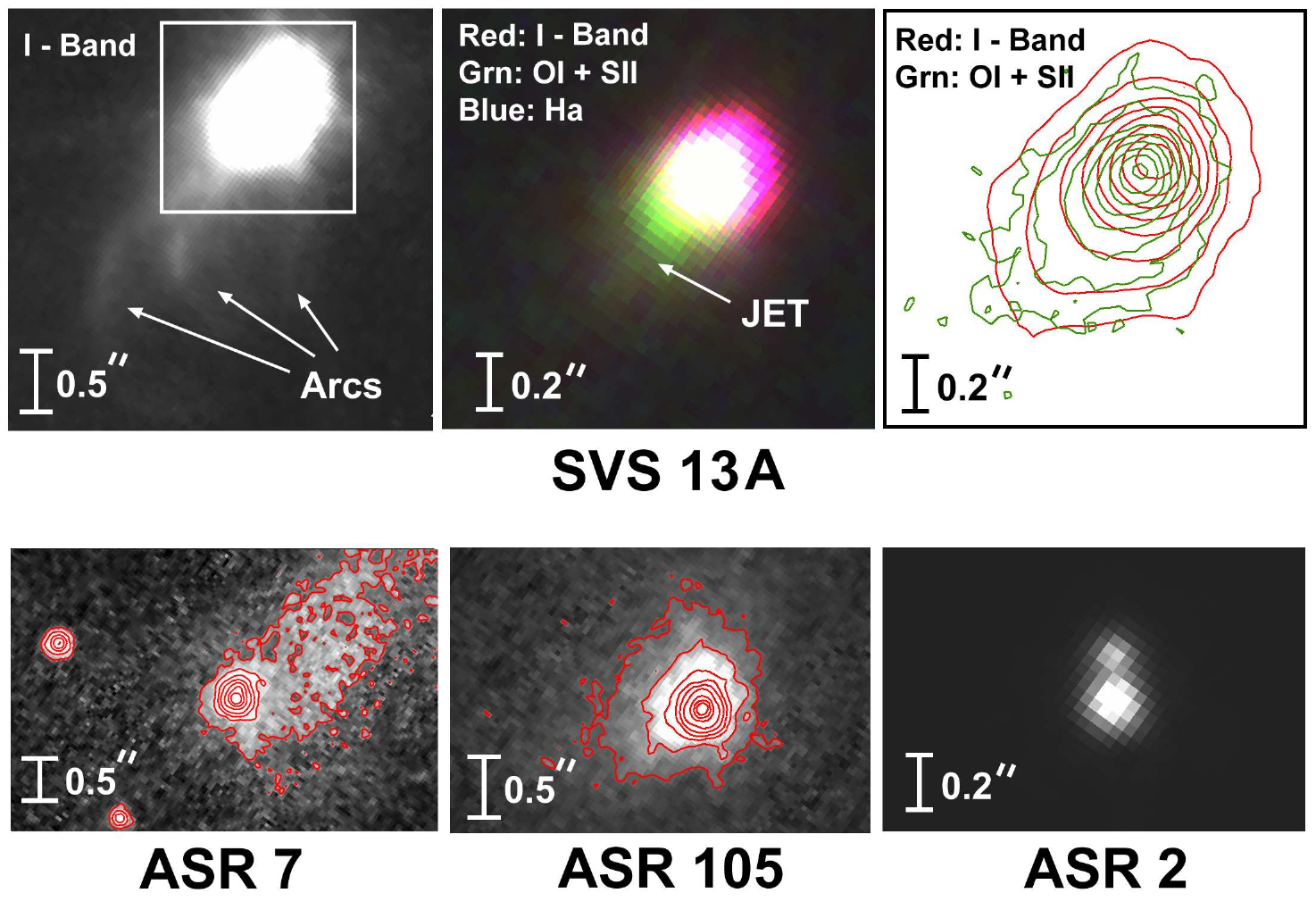}
\caption{Compilation of stellar objects in the field of view that are
not simple point sources. North is up and east to the left in all images
and adjacent contours are a factor of two. The scale bars of
0.2$^{\prime\prime}$ and 0.5$^{\prime\prime}$ correspond to 60~AU and 300~AU,
respectively.  Top: SVS~13A, the driving source of the HH~7-11 flow, 
has several curved arcs in the I-band image and a microjet visible as an 
extension to the southeast of the source in the [O~I] + [S~II] composite. 
Bottom: Both ASR 105 and ASR 7 show extended cavities in their I-band images,
while ASR~2 is a close binary in all the narrowband and
broadband images (F656N shown here).
}
\label{fig:stars}
\end{figure}

\begin{figure}
\centering
\includegraphics[angle=0,scale=1.00,width=5.3in]{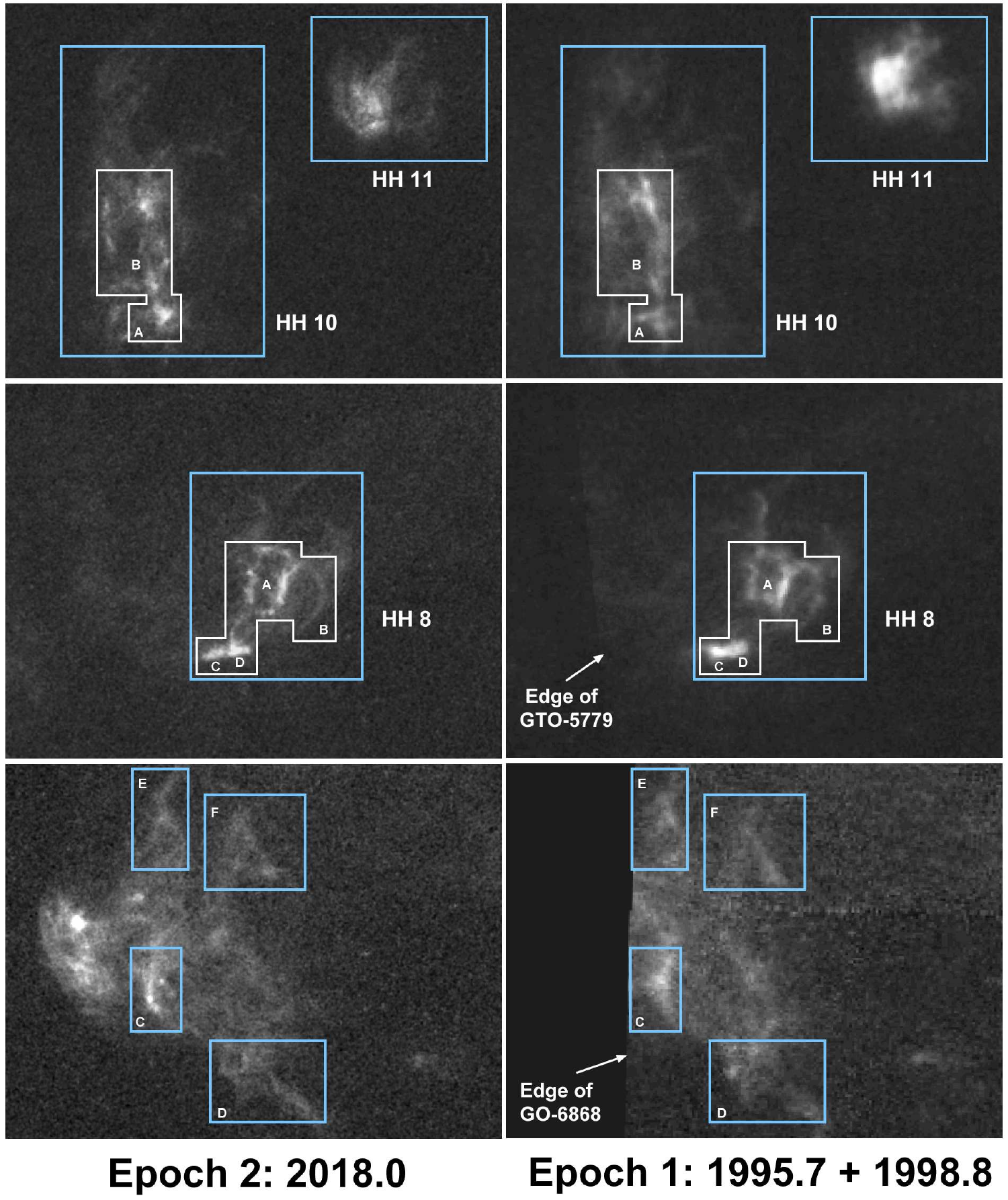}
\caption{Proper Motions and Variability in HH~7-11. The images are H$\alpha$,
and combine GTO-5779 and GO-6868 to produce epoch 1. 
Our new images are shown at left as epoch 2. White boxes in the images
are shown for reference purposes only, and are fixed in the sky.
These boxes highlight secular variability 
if the reader glances quickly between the left and right
panels. The light-blue boxes, also fixed regions of the sky, 
are used for the proper motion measurements in Table~1.
}
\label{fig:time}
\end{figure}

\begin{figure}
\centering
\includegraphics[angle=0,scale=1.00]{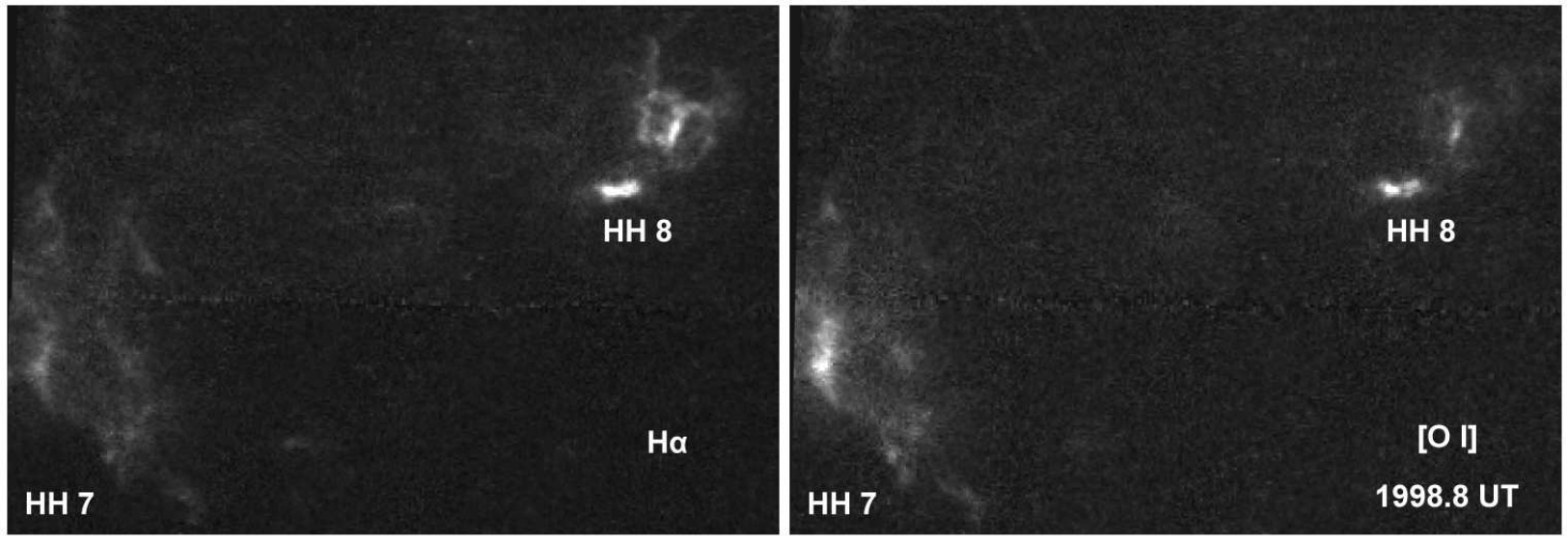}
\caption{Proper Motions and Variability in HH~7 and HH~8. An animated version
of this figure is available. In the animation, the
images for H$\alpha$ (left) and [O~I] (right) switch between epoch 1
and epoch2.
}
\label{fig:movie7}
\end{figure}

\begin{figure}
\centering
\includegraphics[angle=0,scale=1.00]{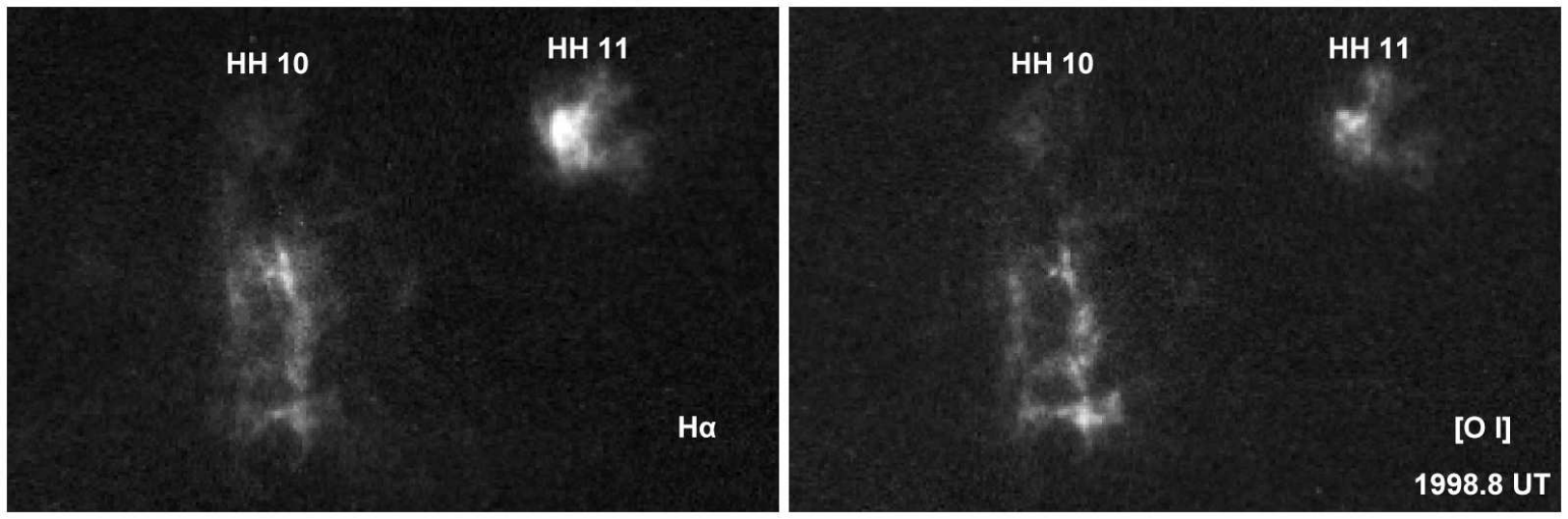}
\caption{Same as Fig.~\ref{fig:movie7} but for HH~10 and HH~11. In the animated
version, H$\alpha$ (left) and [O~I] (right) alternate between epoch 1
and epoch2.
}
\label{fig:movie11}
\end{figure}

The HH~7 bow shock is arguably the best object in the sky for studying 
how molecular emission within C-shocks relates to its
J-shock counterpart traced by optical lines. 
Figs.~\ref{fig:hh7c} and \ref{fig:working-surface} reveal that the brightest portion of HH~7A
has an unusual serrated appearance. We take this feature to be
the working surface, where clumps in the jet 
decelerate as they encounter the dense ambient cloud
(see Fig.~\ref{fig:working-surface}). The brightest knot splits into three smaller
components down near the spatial resolution limit of the observations. These bright knots
are trailed by what appears to be a wake. Apparently parts of the
jet become small, bullet-like clumps before they decelerate in the HH~7 working surface.
The wavy morphology of the working surface could give rise to
Mach stems where curved shocks intersect \citep{hartigan16}, although any such structures 
would be unresolved in these data.
The feature marked as `Inner Rim' seems to mark the edge leading bow shock, as it 
encloses the extent of the H$\alpha$ and forbidden line emission. 
Unfortunately proper motion measurements are not yet available
for this region owing to the pointing errors in the early HST images. 
Interestingly, our I-band image also shows that an `Outer Rim' cavity extends to the east
beyond the forbidden line emission. How this cavity relates to the H$_2$ emission
is unclear, as existing ground-based H$_2$ images do not locate the molecular gas with
enough precision.

Because they represent distinct structures of moving gas, we
can use both HH~7C and HH~11 to measure flow velocities, orientation angles, and 
dynamical ages for these objects. 
Emission from HH~7 occurs primarily along a series of three nested
arcs that resemble bow shocks, HH 7A, 7B, and 7C (Sec.~\ref{sec:ratios}), and we have
proper motion measurements for HH~7C (Table 1).  Combining this measurement
with the observed radial velocity at
this location of $-$82 km$\,$s$^{-1}$ with respect to the molecular cloud \citep{sb87},
we obtain a flow velocity of 90 km$\,$s$^{-1}$ and 
an orientation angle of 25 $\pm$ 5 degrees from the line of sight for HH~7C.
Similarly, using a radial velocity of $-$200 km$\,$s$^{-1}$ for HH~11 we find a flow
velocity of 220 km$\,$s$^{-1}$, and an orientation angle of 24 $\pm$ 2 degrees.
However, even though their orientation angles to the line of sight are essentially
the same, the velocity vectors of HH~11 and HH~7 are not parallel to one another
because the jet is deflected to the east by HH~8 and HH~10 (see below).
The dynamical ages are 2360 years and 285 years, respectively, for HH~7C and HH~11.
Continuing on its current trajectory, HH~11 should overrun HH~10 around the year
2180, and produce a strong shock wave at that time.

\begin{figure}
\centering
\includegraphics[angle=0,scale=1.00]{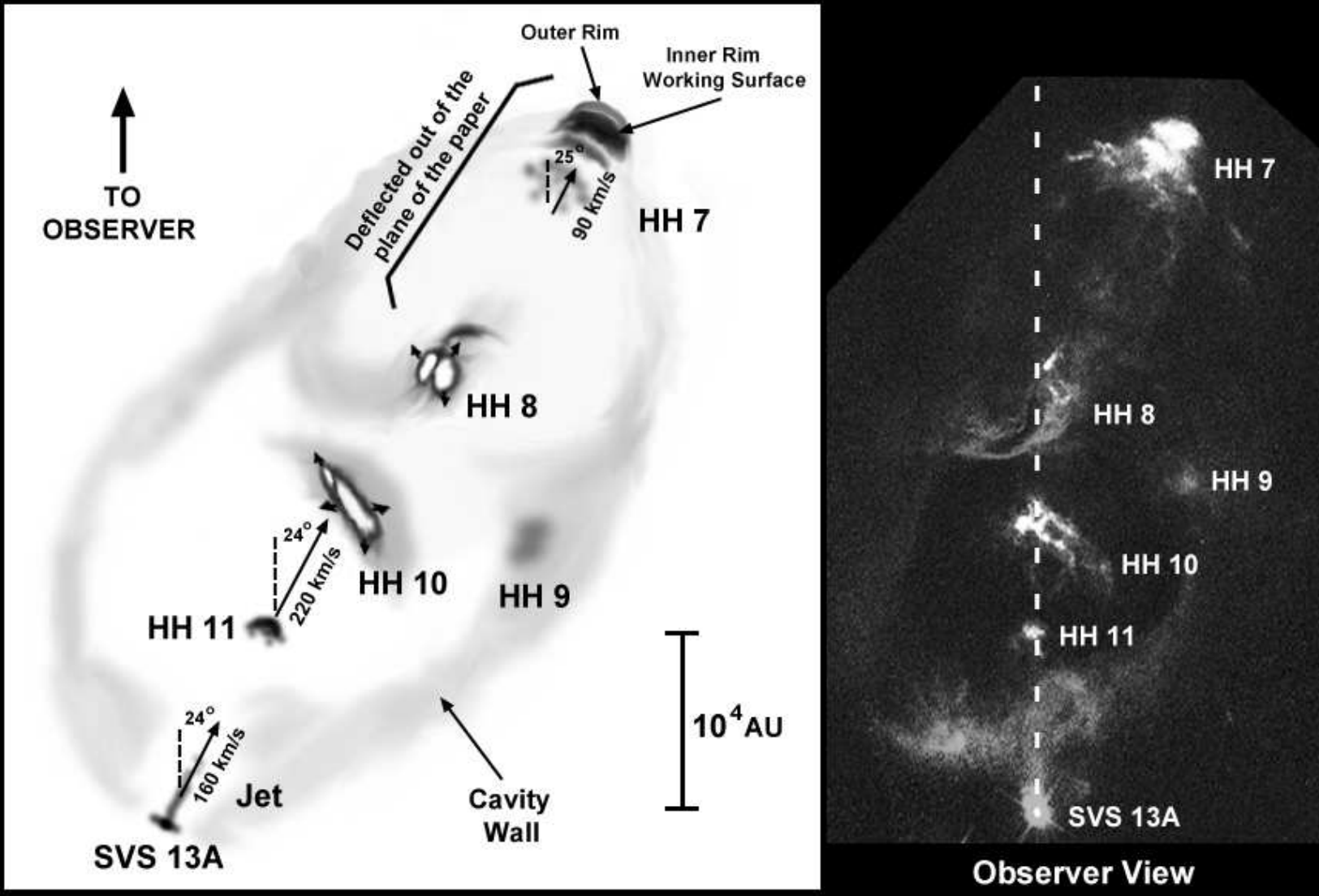}
\caption{Schematic of the HH~7-11 Outflow. Left: HH~11 moves in the plane of
the diagram at an angle of $\sim$ 24 degrees to the line of sight. HH~8 and HH~10 are
variable structures that trace where the jet interacts with 
obstacles along its path, and HH~7 is the terminal
bow shock. Jet gas incident to HH~7 also moves at $\sim$ 25 degrees to the line
of sight, but is tilted out of the plane of the paper, accounting for a shift
of 55 degrees in the direction of proper motion relative to that of HH~11. Right:
Greyscale image of Fig.~\ref{fig:overview}, showing the outflow from the observer's
reference frame. 
}
\label{fig:cartoon}
\end{figure}

Fig.~\ref{fig:cartoon} summarizes the proper motion results.  The jet currently emerges with a
space velocity of $\sim$ 160 km$\,$s$^{-1}$ at PA $\sim$ 145 degrees. We take its
orientation to the line of sight to be 24 degrees, the same as that of HH~11.
Some 12,600~AU (5100~AU projected) along the jet we encounter HH~11, a bullet-like ejection with a velocity
of 220 km$\,$s$^{-1}$ and an inclination of 24 degrees. Its shock velocity is
less than 90 km$\,$s$^{-1}$ owing to the lack of [O~III], but has a line width of
$\sim$ 70 km$\,$s$^{-1}$ \cite{sb87}. Taking this width as an indication of
the shock velocity \citep{hrh87}, the jet material ahead of HH~11 moves at
$\sim$ 150 km$\,$s$^{-1}$ away from the source. The fading of HH~11 could arise
from a lower density, or a faster preshock medium (lower shock velocity)
in the second epoch relative to that of the first epoch.

The jet disappears between HH 11 and HH~7, where we see only the clumpy, variable
stationary objects HH~8 and HH~10 as shocked knots along the flow's periphery.
By the time we see clear evidence for jet material again in HH~7C,
the orientation to the line of sight is essentially the same as it was for HH~11,
{\it but the proper motion vector has shifted by about 55 degrees towards the east.}
As depicted in Fig.~\ref{fig:cartoon}, because
the flow has a significant component towards the observer, a small
deflection of the jet can produce a larger angular change in the proper motion vector.
In this case, it is possible to change the proper motion vector orientation
by 55 degrees by redirecting the jet $\sim$ 23 degrees, something 
easy to do for a 150 km$\,$s$^{-1}$ (Mach $\sim$ 15) jet via an oblique
shock. The only evidence that the jet continues to flow through
HH~8 and HH~10 is that the radial velocities of these objects are substantially
blueshifted ($\gtrsim$ 50 km$\,$s$^{-1}$) relative to the cloud.
Whether or not the jet forms a continuous stream, its velocity
has declined to$\sim$ 90 km$\,$s$^{-1}$ at HH~7C. At this point the jet
seems to become more fragmentary, breaking into smaller clumps that enter the working
surface as small bullets in HH~7A (Fig.~\ref{fig:working-surface}).

The HH~7 bow shock could be more or less stationary, with
a very slow shock propagating into the molecular gas, but we do not yet have
proper motion data to test this idea.
Given a typical evolutionary timescale of $10^5$ $-$ $10^6$ years for the protostar,
it is highly unlikely that we happen to be observing this jet when it
is only a few thousand years old. Instead, if the jet encounters a
much denser molecular cloud it will pile up into the shell that
defines HH~7. Unable to break out of this cocoon, the jet nevertheless
creates a cavity and gradually drives a molecular flow along the periphery
of this shell. Normally we think of stellar jets as being much denser than
their surroundings and penetrating for large distances. However, the HH~7-11 outflow
seems to be an exception to that rule. The available data indicate that the
flow is very young and has yet to fully emerge from the molecular cloud core that created
the driving source SVS~13A.

The forbidden line emission here all aligns nearly along the axis of a reflected-light cavity whose base is
located at SVS~13A, itself an embedded IR source with a disk that drives a
microjet down the middle of the cavity. These facts argue strongly that all of the line emission
in HH~7-11 originates from a single outflow. However, a dark hole in the reflected light
cavity surrounds HH~11 and HH~10 (Fig.~\ref{fig:overview}) and this striking feature
remains unexplained in the above scenario. We initially considered a model
where HH~7 and HH~11 arose from different flows projected onto
nearly the same line of sight, perhaps from the two components of the SVS~13 binary
\citep{plunkett13}. This dual jet model could explain the dark hole by having the 
HH~11 jet oriented more towards the observer so it 
punches through the larger cavity created by the HH~7 jet.
However, this model is not consistent with the proper motion measurements, 
which show similar inclinations for HH~7C and HH~11 relative
to the observer.  It is worth noting that this region does contain
several outflows \citep[e.g.][]{dionatos17}, including a second molecular flow
detected at mm-wavelengths along the western edge of the HH~7-11 cavity \citep{lef17}.
However, none of these other flows seem to affect the shocked optical emission
in this region.

\section{Summary}
\label{sec:summary}

When combined with archival images taken 20 years ago, the new
narrowband emission line images of the HH~7-11 stellar jet reported in
this paper allow us to measure precise proper motions throughout the flow
for the first time.  Our program is also the first to use a quad filter
on HST to isolate only [S~II] $\lambda$6716, making it possible to measure electron
densities and line excitations everywhere throughout the flow with
spatial resolution of $\sim$ 20~AU.  The new images also uncovered several
reflection nebulae suggestive of disks and cavities around embedded yonug stars
in the area, and resolved one such object as a subarcsecond binary.

The source of the HH 7-11 outflow is SVS~13A, which drives a
jet down the axis of a reflected-light cavity centered on the HH objects.
The images have uncovered a new morphological structure not seen
clearly before in any HH jet - both HH~8 and HH~10 are variable, ring-like
structures that have no bulk proper motions but 
the shocked gas has a significant radial velocity. 
In HH~8 the rings appear to be expanding, while the motions in HH~10 are
more chaotic.  We interpret these features to be locations where the jet
deflects from an obstacle along its path. In this sense, the HH~7-11 jet
is similar to HH~110, which also exhibits line emission as the jet
deflects from an obstacle in the flow \citep{riera03,lopez05,hartigan09}.
The two obvious bow shocks, HH~7 and HH~11, have similar orientation
angles of $\sim$ 25 degrees to the line of sight.
HH~11 appears as a typical HH bow shock with a substantial
proper motion, and is fading with time, while HH~7 consists of multiple
arcs and knots.

The HH~7 bow shock seems to mark the end of the flow,
though no proper motions yet exist for the apex of this feature.
A complex working surface defines the region where knots in
the fragmentary jet decelerate.
The bow shock in HH~7 also emits strongly in H$_2$, making it
an ideal target to study how C-shocks connect to their J-shock
counterparts.  Such a comparison requires requires high-spatial
resolution observations of the shocked H$_2$ emission, a high
priority for JWST once it becomes operational.

The presence of emission line structures that align with the jet but do
not participate in the outflow reinforces the need for caution when
attempting to convert line luminosities in jets to mass outflow rates.
Any such attempts should carefully consider how entrained material and
shocks within ambient gas affect the mass loss estimates. These estimates
are likely be reliable only close to the source before multiple shock
waves alter the geometry of the flow.

\acknowledgements

Support for Program number GO-15257 was provided by NASA through a grant from the
Space Telescope Science Institute, which is operated by the Association of
Universities for Research in Astronomy, Incorporated, under NASA contract NAS5-26555.
This work has made use of data from the European Space Agency (ESA) mission
Gaia (https://www.cosmos.esa.int/gaia), processed by the Gaia Data Processing
and Analysis Consortium (DPAC, https://www.cosmos.esa.int/web/gaia/dpac/consortium).
Funding for the DPAC has been provided by national institutions, in particular
the institutions participating in the Gaia Multilateral Agreement.

\null
\begin{center}
\begin{deluxetable}{cccccc}
\singlespace
\tablenum{1}
\tablewidth{0pt}
\tablecolumns{5}
\tabcolsep = 0.06in
\parindent=0em
\tablecaption{Proper Motions in the HH 7-11 Jet}
\startdata
\noalign{\medskip}
\noalign{\medskip}
\noalign{\hrule}
\noalign{\medskip}
Object$^{\rm a}$&
$\mu_{\rm RA}$$^{\rm b}$&
$\mu_{\rm DEC}$$^{\rm c}$&
$\mu$ (km$\,$s$^{-1}$)$^{\rm d}$\\
\noalign{\smallskip}
\noalign{\hrule}
\noalign{\smallskip}
HH 7C & 2.7 $\pm$ 0.4& 0.1 $\pm$ 0.4& 38 $\pm$ 7 \\
HH 7D & 0.9 $\pm$ 0.4& 0.4 $\pm$ 0.4& 14 $\pm$ 7 \\
HH 7E & 0.5 $\pm$ 0.4& $-$0.7 $\pm$ 0.4 & 12 $\pm$ 7\\
HH 7F & 1.2 $\pm$ 0.4& 0.6 $\pm$ 0.4 & 19 $\pm$ 7\\
HH 8  & 0.2  $\pm$ 0.3& 0.1 $\pm$ 0.3 & $<$ 6\\
HH 10 & 0.2  $\pm$ 0.3& $-$0.2 $\pm$ 0.3 & $<$ 6\\
HH 11 & 3.9 $\pm$ 0.3& $-$5.0 $\pm$ 0.3 & 91 $\pm$ 6\\
\enddata
\tablenotetext{a:}{Objects defined by blue boxes in Fig.\ref{fig:time} }
\tablenotetext{b:}{Proper motion in RA (arcsec/century)}
\tablenotetext{c:}{Proper motion in DEC (arcsec/century)}
\tablenotetext{d:}{Tangential velocities, assuming a distance of 300~pc}
\end{deluxetable}
\label{table1}
\end{center}
 

\clearpage
\begin{thebibliography}{}

\bibitem[Anglada et~al.(2000)]{anglada00} 
Anglada, G., Rodr\'iguez, L.F., \& Torrelles, J.M. 2000 ApJ 542, L123

\bibitem[Aspin et~al.(1994)]{aspin94} 
Aspin, C., Sandell, G., \& Russell, A.P.G. 1994, A\&AS 106, 165

\bibitem[Belikov et~al.(2002)]{belikov02} 
Belikov, A., Kharchenko, N., Piskunov, A., Schilbach, E., \& Scholz, R.-D. 
2002, A\&A, 387, 117

\bibitem[B\"ohm et~al.(1983)]{bohm83} 
B\"ohm, K.-H., Brugel, E.W., \& Olmsted, E. 1983, A\&A 125, 23

\bibitem[de Colle et~al.(2010)]{deC10} 
de Colle, F., del Burgo, C. \& Raga, A.C.  2010 ApJ 721, 929

\bibitem[de Zeeuw et~al.(1999)]{deZ99} 
de Zeeuw, P.T., Hoogerwerf, R., Bruijne, J.H.J., Brown, A.G.A., \& Blaauw, A.
1999, AJ 117, 354

\bibitem[Dhabal et~al.(2018)]{herschel} 
Dhabal, A., Mundy, L.G., Rizzo, M.J., Storm, S., \& Teuben, P.  2018, ApJ 853, 169

\bibitem[Dionatos \& G\"udel(2017)]{dionatos17} 
Dionatos, O., \& G\"udel, M. 2017, A\&A 597, A64

\bibitem[Dopita \& Sutherland(2017)]{ds17} 
Dopita, M. \& Sutherland, R.  2017 ApJS 229, 35 

\bibitem[Dressel(2019)]{dressel19} 
Dressel, L. 2019, {\it Wide Field Camera 3 Instrument Handbook, Version 11.0}, (Baltimore: STScI)

\bibitem[Foster et~al.(2015)]{foster15} 
Foster, J.B, et al. 2015, ApJ 799, 136

\bibitem[Frank(2007)]{frank07} 
Frank, A. 2007, Ap\&SS 307, 35

\bibitem[Frank et~al.(2014)]{frank14} 
Frank, A. et~al.  in {\it Protostars and Planets VI}, H. Beuther,
R.S. Klessen, C.P. Dullemond, \& T.K. Henning eds.,(Tucson:U of A Press), p451

\bibitem[Gaia Collaboration (2016)]{gaia1} 
Gaia Collaboration, 2016, A\&A 595, 1

\bibitem[Gaia Collaboration (2018)]{gaia2} 
Gaia Collaboration, 2018, A\&A 616, 1

\bibitem[Garden et~al.(1990)]{garden90} 
Garden R. P., Russell A. P. G. \& Burton M. G. 1990, ApJ 354 232

\bibitem[Geballe et~al.(2017)]{geballe17} 
Geballe, T., Burton, M.E., \& Pike, R.E. 2017 ApJ 837, 83

\bibitem[Gennaro et~al.(2018)]{gennaro18} 
Gennaro, M. et al. 2018, {\it WFC Data Handbook}, Version 4.0, (Baltimore:STScI)

\bibitem[Greissl et~al.(2007)]{greissl07} 
Greissl, J., Meyer, M.R., Wilking, B.A., Fanetti, T., Schneider, G.,
Greene, T.P., \& Young, E. 2007, AJ 133, 1321

\bibitem[Hartigan et~al.(2009)]{hartigan09} 
Hartigan, P., Foster, J., Wilde, B., Coker, R., Rosen, P., Hansen, J., Blue,
B., Williams, R., Carver, R., \& Frank, A.  2009, ApJ 705, 1073

\bibitem[Hartigan et~al.(2011)]{hartigan11} 
Hartigan, P., Frank, A., Foster, J.M., Wilde, B.H., Douglas, M., Rosen, P.A.,
Coker, R.F., Blue, B.E., \& Hansen, J.F.  2011, ApJ 736, 29 

\bibitem[Hartigan et~al.(2016)]{hartigan16} 
Hartigan, P., Foster, J., Frank, A., et~al. 2016, ApJ 823, 148

\bibitem[Hartigan \& Morse(2007)]{hm07} 
Hartigan, P. \& Morse, J. 2007, ApJ 660, 426

\bibitem[Hartigan et~al.(1994)]{hrm94} 
Hartigan, P., Morse, J., \& Raymond, J. 1994, ApJ 436, 125

\bibitem[Hartigan et~al.(1987)]{hrh87} 
Hartigan, P., Raymond, J., \& Hartmann, L. 1987, ApJ 316, 323

\bibitem[Hartigan \& Wright(2015)]{hw15} 
Hartigan, P. \& Wright, A. 2015, ApJ 811, 12

\bibitem[Heathcote et~al.(1996)]{heathcote96} 
Heathcote, S., Morse, J., Hartigan, P., Reipurth, B., Schwartz, R.D.,
Bally, J., \& Stone, J.  1996, AJ 112, 1141

\bibitem[Herbig(1974)]{herbig74} 
Herbig, G. 1974, Lick Obs. Bull. No. 658

\bibitem[Herbig \& Jones(1983)]{hj83} 
Herbig, G.H. \& Jones, B.F. 1983, AJ 88, 1040

\bibitem[Hodapp \& Chini(2014)]{hodapp14} 
Hodapp, K.W., \& Chini, R. 2014, ApJ 794, 169

\bibitem[Khanzadyan et~al.(2003)]{khan03}
Khanzadyan T., Smith M. D., Davis C. J., Gredel, R.,
Stanke, T., \& Chrysostomou, A. 2003, MNRAS 338 57

\bibitem[Lef\'evre et~al.(2017)]{lef17} 
Lef\'evre, C., Cabrit, S., Maury, A.J. et~al. 2017, A\&A 604, L1

\bibitem[Lopez et~al.(2005)]{lopez05} 
Lopez, R., Estalella, R., Raga, A., Reira, A., Reipurth, B., \& Heathcote, S.
2005, A\&A, 432, 567

\bibitem[Lucas et~al.(2008)]{lucas08} 
Lucas, P.W., Hoare, M.G., Longmore, A.C., et~al. 2008, MNRAS 391, 136

\bibitem[Nolan et~al.(2017)]{nolan17} 
Nolan, C. A., Salmeron, R., Federrath, C., Bicknell, G. V., \& Sutherland, R. S.
2017, MNRAS 471, 1488

\bibitem[Noreiega-Crespo \& Garnavich(2001)]{noriega01} 
Noreiega-Crespo A., \& Garnavich, P.M. 2001 AJ 122, 3317

\bibitem[Oasa et~al.(2008)]{oasa08} 
Oasa, Y., Tamura, M., Sunada, K., \& Sugitani, K. 2008, AJ 136, 1372

\bibitem[Pike et~al.(2016)]{pike16} 
Pike, R.E., Geballe, T.R., Burton, M.G., \& Chrysostomou, A. 2016, ApJ 822, 82

\bibitem[Plunkett et~al.(2013)]{plunkett13} 
Plunkett, A.L., Arce, H.G., Corder, S.A.,
Mardones, D., Sargent, A.I., \& Schnee, S.L. 2013, ApJ 774, 22

\bibitem[Raga et~al.(1996)]{raga96} 
Raga, A.C, Canto, J., \& Steffen, W. 1996, QJRAS 37, 493

\bibitem[Raga et~al.(2016)]{raga16} 
Raga, A.C, Reipurth, B., Vel\'azquez, P.F., Esquivel, A., \& Bally, J. 2016, AJ 152, 186

\bibitem[Raga et~al.(2012)]{raga12} 
Raga, A.C, Noriega-Crespo, A., Carey, S.J., \& Arce, H. 2012, AJ 145, 28

\bibitem[Riera et~al.(2003)]{riera03} 
Riera, A., Raga, A., Reipurth, B., Amram, P., Boulesteix, J., Canto, J., \&
Toledano, O. 2003, AJ, 126, 327

\bibitem[Rebull(2015)]{rebull15} 
Rebull, L. 2015, AJ 150, 17

\bibitem[Reipurth et~al.(1997)]{reipurth97} 
Reipurth, B., Bally, J., \& Devine, D. 1997, AJ 114, 2708

\bibitem[Romero et~al.(2017)]{romero17} 
Romero, G.E., Boettcher, M., Markoff, S., \& Tavecchio, F. 2017, Sp.Sci.Rev. 207, 5

\bibitem[Snell \& Edwards(1981)]{se81} 
Snell, R. \& Edwards, S. 1981, ApJ 251, 103

\bibitem[Solf  \& B\"ohm(1987)]{sb87} 
Solf, J., \& B\"ohm, K.-H. 1987, AJ 93, 1172

\bibitem[Solf \& B\"ohm(1990)]{sb90} 
Solf, J., \& B\"ohm, K.-H. 1990, ApJ 348, 297

\bibitem[Shaw \& Dufour(1994)]{sd94} 
Shaw, R.A. \& Dufour, R. 1994, ASPC 61, 327

\bibitem[Strom et~al.(1976)]{svs76} 
Strom, S.E., Vrba, F.J., \& Strom, K.M. 1976, AJ 81, 314

\bibitem[Williams(1973)]{williams73} 
Williams, R.E. 1973, MNRAS 164, 111

\bibitem[Winston et~al.(2010)]{winston10} 
Winston, E., Megeath, S.T., Wolk, S.J. et~al. 2010, AJ 140, 266

\bibitem[Yuan \& Neufeld(2011)]{yuan11} 
Yuan, Y., \& Neufeld, D.A. 2011, ApJ 726, 76

\bibitem[Yuan et~al.(2012)]{yuan12} 
Yuan, Y., Neufeld, D.A., Sonnentrucker, P., Melnick, G.J., \& Watson, D.M.
2012, ApJ 753, 126

\bibitem[Zanni \& Ferreira(2013)]{zanni13} 
Zanni, C., \& Ferreira, J. 2013, A\&A 550, 99

\end{thebibliography}
\end{document}